\documentclass[aps,prb,amsmath,showpacs,amssymb]{revtex4}
\usepackage{graphicx}
\usepackage{color}
\begin{document}
\draft
\title{Defect capturing and charging dynamics and their effects on magneto-transport of electrons in quantum wells}

\author{Andrii Iurov\footnote{E-mail contact: aiurov@mec.cuny.edu, theorist.physics@gmail.com
},$^{1}$ Danhong Huang\footnote{E-mail contact: danhonghuang1647@outlook.com},$^{2}$ Godfrey Gumbs,$^{3}$ Paula Fekete,$^{4}$ and Fei Gao$^{5}$}
\affiliation{
$^{1}$Department of Physics and Computer Science, Medgar Evers College of the City University of New York, 
Brooklyn, NY 11225, USA\\
$^{2}$Air Force Research Laboratory, Space Vehicles Directorate, Kirtland Air Force Base, NM 87117, USA\\
$^{3}$Department of Physics and Astronomy, Hunter College of the City University of New York, 695 Park Avenue, New York, NY 10065, USA\\
$^{4}$Department of Physics and Nuclear Engineering, US Military Academy at West Point, West Point, New York 10996, USA\\
$^{5}$Department of Nuclear Engineering and Radiological Sciences, University of Michigan, Ann Arbor, MI 48109, USA
}

\date{\today}

\begin{abstract}
The defect corrections to polarization and dielectric functions of Bloch electrons in quantum wells are first calculated. Following this, we derive the first two moment equations from  Boltzmann transport theory and apply them to explore defect effects on magneto-transport of Bloch electrons. Meanwhile, we obtain analytically the momentum-relaxation time and mobility tensor for Bloch electrons making use of the screened defect-corrected polarization function. Based on quantum-statistical theory, we further investigate the defect capture and charging dynamics by employing a parameterized physics model for defects to obtain defect wave functions. After this, both capture and relaxation rates, as well as density for captured Bloch electrons, are calculated self-consistently as functions of temperature, doping density and different  defect types. By applying the energy-balance equation, the number of occupied energy levels and chemical potential of defects are determined, with which the transition rate for defect capturing is obtained. By using these results, defect energy-relaxation, capture and escape rates, and Bloch-electron chemical potential are obtained self-consistently. At the same time, the Bloch-electron energy- and momentum-relaxation rates, as well as the current suppression factor, are also investigated quantitatively. Finally, by combining all these studies together, the temperature dependence of the Hall and longitudinal mobilities is demonstrated for Bloch electrons in either single- or multi-quantum wells, which can be utilized for quantifying burst noise in transistors and blinking noise in photo-detectors.
\end{abstract}
\maketitle

\section{Introduction}
\label{sec-1}

Point defects (vacancies, interstitials and voids) are generated by displacements of atoms from their thermal-equilibrium lattice sites,\,\cite{book-gary,book-sigmund}
where the lattice-atom displacements are mainly caused by a proton-irradiation-induced
primary knock-on atom (PKA) on a time scale shorter than $100$\,ps for building up point defects without reactions.
These initial displacements are followed immediately by defect mutual recombinations or reactions with sinks
(clustering or dissolution of clusters for point-defect stabilizations)\,\cite{add2,gao6} on a time scale shorter than $10\,$ns,
then possibly by thermally-activated defect migrations\,\cite{gao5} up to a time scale much longer than $10$\,ns (steady-state distributions).
Such atom displacements depend not only on the energy-dependent flux of protons but also on the differential energy transfer cross sections (probabilities) for collision between
atoms, interatomic Coulomb interactions and even kinetic-energy loss to core-level electrons of an atom (ionizations). The sample temperature at which the irradiation has been performed
also significantly affects the diffusion of generated defects,
their stability as clusters and the formation of Frenkel pairs.\,\cite{gao7}
One of the effective calculation methods for studying the non-thermal spatial-temporal distributions of proton-irradiation induced point defects is the molecular-dynamics (MD) model\,\cite{gao2}. However,
the simulation-system size increases quadratically with the initial kinetic energy of protons and the time scale can easily run up to several hundred
picoseconds. In this case, the defect reaction process driven by thermal migration cannot be included in the MD model due to its much longer time scale. Practically,
if the system time evolution goes beyond $100$\,ps, either the kinetic lattice Monte-Carlo\,\cite{gao1} or the diffusion-reaction equation\,\cite{ryazanov,stoller} method should be employed instead.
\medskip

Experimentally, positron annihilation spectroscopy is found very useful for investigating vacancy-defect properties in semiconductors.\,\cite{new1} By combining experimental and theoretical approaches together, it becomes possible to identify different species of defects and their chemical surroundings, including charge states and defect levels within the semiconductor bandgap. Theoretically, point defects and impurities can strongly affect both optical and transport properties of semiconductor materials\,\cite{feigao} and play a key role 
on their performance in device applications.\,\cite{dft-1} Very importantly, first-principles density functional computations are recognized as a powerful approach that complements experiments and is enable to serve as a predictive tool in the identifying and characterizing defect electronic states.\,\cite{dft-2,new3} From application perspective, however, only very few of microscopic-level theories have been proposed for systematically describing dynamics of defect generation,\,\cite{new6,new7,add-4}, especially the dynamics for defect effects\,\cite{feigao,ajss} with a quantum-statistical theory, including transition rate, charging, capture, relaxation, scattering and current suppression.
\medskip

In the presence of defects, dangling bonds attached to these point defects can capture Bloch electrons through multi-phonon emission to form localized charge centers.
These randomly-distributed charge centers will further affect electron responses to either an external ballistic electron beam\,\cite{book} or incident photons\,\cite{add-1}.
Physically, the defect modifications to the electron response function can be included by a vertex correction\,\cite{add-1} to an unscreened electron polarization function under the ladder approximation\,\cite{stern} (LA).
Additionally, the many-body screening by other electrons in a quantum well can be included by employing the random-phase approximation\,\cite{stern,add-3} (RPA).
The quantum-statistical theory presented in this paper is crucial for fully understanding the mechanism in characterizing defects,\,\cite{hallen} defect effects,\,\cite{doan}
as well as for developing an effective mitigation scheme in early design stages of electronic devices. Aided by this multi-timescale microscopic theory,\,\cite{ajss} the experimental characterization of post-irradiated
test devices\,\cite{hubbs} is able to provide more useful information on the device architecture's susceptibility to space radiation effects\,\cite{simoen}. Furthermore, our quantum-statistical physics model should also allow for accurate
prediction of device-performance degradation in combination with the
space weather forecast model\,\cite{swf,cooke} for a particular satellite orbit. With this paper, we expect to bridge the gap between researchers studying radiation-induced damage in materials\,\cite{book-gary,book-sigmund,add-4,add-5}
and others characterizing irradiation induced device-performance degradation.\,\cite{vince,chris}
Finally, by studying fast defect capturing and charging dynamics, the numerical results for both capture and relaxation times of a point defect,
as well as captured electron density, can be obtained as functions of lattice temperature, electron density and various types of defects and applied to device simulations for analyzing burst noise in transistors, and meanwhile, blinking noise, 
reduced quantum efficiency and increased dark current in photo-detectors.
\medskip

As for applications of the current theory, we would like to emphasize that space-based infrared (IR) imaging is expected to face the most stringent performance requirements on the quantum-well (QW) focal plane array (FPA)\,\cite{fpa,fpa2} due to the high payload cost, complexity, and remoteness the space environment imparts as well as the order of magnitude or more lower incident thermal-radiation photon-flux levels that occur in space environments compared to terrestrial ones. Space applications also set up the unique requirement for radiation tolerance (or rad-hardness) on two elements of the hybridized QW-FPA, i.e., the Si-CMOS (complementary metal-oxide-semiconductor) read-out integrated circuit and the photodetector array. FPAs developed for these purposes in a space environment are meticulously characterized for their sensitivity, uniformity, operability and radiation hardness. IR detector arrays operated in the space environment are subjected to a variety of radiation sources while in orbit, e.g., electrons, protons and some heavy ions confined by the earth's magnetic field (Van Allen radiation belts). This indicates that QW photodetectors for space-based surveillance or space-situational awareness must be characterized in advance and should acquire not only high performance (high quantum efficiency and low dark current density)\,\cite{fpa3,fpa4} but also radiation tolerance or ability to withstand the effects of the radiation they would expect to encounter on a given orbit. Detector technologies that operate in the harsh radiation environment of space with better radiation tolerance would lead to greater flexibility in orbit selection, technical applications and system sustainability, and therefore, are of more value to the space-based sensing community.
\medskip

The rest of the paper is organized as follows. In Sec.\,\ref{sec-2}, we present our theoretical model 
to investigate the point-defect effects on the polarization and dielectric functions of electrons in a quantum well.
We acquire the first two moments from Boltzmann transport equation in Sec.\,\ref{sec-3} for revealing the role of defects played in quantum-well magneto-transport.
In Sec.\,\ref{sec-4}, we calculate the momentum-relaxation time and mobility tensor for electron magneto-transport based on our obtained defect-corrected polarization function.
By going beyond a simple $\delta$-function for point-defect probability function, we study in Sec.\,\ref{sec-5} the capture and charging dynamics of defects using quantum-statistical theory and obtain
the capture and relaxation rates as well as the density of captured subband electrons.
Finally, a brief summary is given in Sec.\,\ref{sec-6}.

\section{Defect Effect on Polarization Function}
\label{sec-2}

Since the wave functions of individual defects are spatially localized, we anticipate that the interaction between electrons and charged defects
can affect mainly the screening to the intralayer Coulomb interaction between electrons.
Therefore, we will start with defect effects in a single quantum well. The theoretical study on
exchange-interaction induced vertex correction to a bare polarization function of electrons in a quantum well has already been reported\,\cite{add-1}
within the ladder approximation. Very recently, the defect-induced vertex correction to a bare polarization function of electrons in multi-quantum wells has also been studied\,\cite{feigao}.
\medskip

For a single narrow quantum well, the separation between two adjacent conduction subbands is very large, and thus only the lowest subband will be occupied by electrons (i.e., {\em electric-quantum limit}).
In this case, the screened electron polarization function $\chi_1(q_\|,\omega)$
can be calculated\,\cite{add-11} through an inverse dielectric function $\epsilon^{-1}(q_\|,\omega)$, according to\,\cite{book}

\begin{equation}
\chi_1(q_\|,\omega)=\epsilon^{-1}(q_\|,\omega)\,\chi_1^{(0)}(q_\|,\omega)\,\Gamma(q_\|,\omega)\ ,
\label{eqn-1}
\end{equation}
where $\Gamma(q_\|,\omega)$ represents an intra-well defect-vertex correction, which can be determined by Eq.\,\eqref{eqn-9} below,
and the bare polarization function $\chi_1^{(0)}(q_\|,\omega)$ takes the form

\begin{eqnarray}
\nonumber
\chi_1^{(0)}(q_\|,\omega)&=&\frac{1}{2\pi^2}\int\limits_0^\infty dk_\|\,k_\|\int\limits_0^{2\pi} d\phi_{{\bf k}_\|,{\bf q}_\|}\,
\left[f_0(\varepsilon_{k_\|}-u_0)-f_0(\varepsilon_{|{\bf k}_\|+{\bf q}_\||}-u_0)\right]\\
&\times&\left\{{\rm Re}\left[\frac{1}{\hbar\omega-i\gamma_0-\varepsilon_{|{\bf k}_\|+{\bf q}_\||}+\varepsilon_{k_\|}}\right]
+i\pi\,{\cal L}_0\left(\hbar\omega-\varepsilon_{|{\bf k}_\|+{\bf q}_\||}+\varepsilon_{k_\|},\,\gamma_0\right)\right\}\ ,\ \ \ \
\label{eqn-2}
\end{eqnarray}
where ${\cal L}_0(a,b)=(b/\pi)/(a^2+b^2)$ is the Lorentz shape function, 
 $\phi_{{\bf k}_\|,{\bf q}_\|}$ is the angle between wave vectors ${\bf k}_\|$ and ${\bf q}_\|$,
$\gamma_0$ is the energy-level broadening, $\displaystyle{\varepsilon_{k_\|}=\frac{\hbar^2 k_\|^2}{2m^\ast}}$ is the subband energy,
$m^\ast$ is the effective mass,
$f_0(x)=\left[1+\exp\left(x/k_BT\right)\right]^{-1}$ is the Fermi function for thermal-equilibrium electrons, $u_{0}$ and $T$ are the chemical potential (evaluated from the subband edge)
and temperature of electrons (or lattice phonons), respectively.
\medskip

Additionally, for a single occupied subband (electric-quantum limit), the inverse dielectric function $\kappa(q_\|,\omega)$ employed in Eq.\,(\ref{eqn-1}) is simply given by
$\displaystyle{\epsilon^{-1}(q_\|,\omega)=\frac{1}{\epsilon(q_\|,\omega)}}$,
where $\epsilon(q_\|,\omega)$ is the dielectric function of electrons and is calculated\,\cite{add-3}
within the RPA as

\begin{equation}
\epsilon(q_\|,\omega)=1
-\,v_{\rm e}(q_\|)\,\chi_1^{(0)}(q_\|,\omega)\,\Gamma(q_\|,\omega)\ ,
\label{eqn-4}
\end{equation}
and the second term contains the defect-vertex correction $\Gamma(q_\|,\omega)$ to the polarization function.
In Eq.\,(\ref{eqn-4}), $v_{\rm e}(q_\|)$ represents the {\em intra-well} Coulomb interaction between electrons, given by\,\cite{add-12}

\begin{equation}
v_{\rm e}(q_\|)=\frac{e^2}{2\epsilon_0\epsilon_{\rm s}(q_\|+q_{\rm s})}\,\int\limits_{-\infty}^\infty dz
\int\limits_{-\infty}^\infty dz'\,{\cal F}^2_{0}(z)\,
\texttt{e}^{-q_\||z-z'|}\,{\cal F}^2_{0}(z^\prime)\ ,
\label{eqn-5}
\end{equation}
where $\epsilon_{\rm s}$ is the host-material dielectric constant, $\displaystyle{{\cal F}_0(z)=\left(1/\sqrt{\pi}\ell_0\right)^{1/2}\texttt{e}^{-z^2/2\ell_0^2}}$
is the wave function corresponding to the lowest subband, $\ell_0=\sqrt{\hbar/m^*\omega_0}$ represents the half width of the quantum well,
$\hbar\omega_0$ measures the energy separation between the ground and the first subband, and

\begin{equation}
q_{\rm s}=\frac{e^2}{2\pi\epsilon_0\epsilon_{\rm s}}\,\int\limits_0^\infty dk_\|\,k_\|
\left[-\frac{\partial f_0(\varepsilon_{k_\|}-u_0)}{\partial\varepsilon_{k_\|}}\right]\ ,
\label{eqn-6}
\end{equation}
which plays the role of the inverse of static screening length\,\cite{add-12}.
Here, the parabolic-potential model employed for electrons confined in a quantum well
is not only reliable  to model a number of existing realistic situations in semiconductor physics
but also proves helpful for considering a simple mathematical model to reveal the details of an
otherwise complicated theory for effects of defects. 	
Especially, this model has been applied satisfactorily to electrons in narrow quantum
wells with high potential barriers, large level separations and low electron density for which tunneling becomes negligible.
Therefore, the overlap of electron wave functions in adjacent wells is insignificant and therefore ignored.
\medskip

For the defect-vertex correction,\,\cite{add-1} $\Gamma(q_\|,\omega)$, introduced in Eqs.\,(\ref{eqn-1}) and (\ref{eqn-4}),
we find the following self-consistent equation within LA

\begin{eqnarray}
\nonumber
\Gamma(q_\|,\omega)&=&1+\frac{1}{2\pi^2}\int\limits_0^\infty dp_\|\,p_\|\,\chi_1^{(0)}(p_\|,\omega)\,\Gamma(p_\|,\omega)\,
\delta(\varepsilon_{q_\|/2}-\varepsilon_{p_\|/2})\int\limits_{-{\cal L}_0/2}^{{\cal L}_0/2} d\xi\,\rho_{\rm d}(\xi)\\
\nonumber
&\times&\int\limits_0^{\pi} d\phi_{{\bf q}_\|,{\bf p}_\|}\,\left|U_{\rm d}(\left|\mbox{\boldmath$q$}_\|-\mbox{\boldmath$p$}_\|\right|,T,\xi)\right|^2\\
&=&1+\frac{2m^*}{\pi^2\hbar^2}\,\chi_1^{(0)}(q_\|,\omega)\,\Gamma(q_\|,\omega)
\int\limits_{-{\cal L}_0/2}^{{\cal L}_0/2} d\xi\,\rho_{\rm d}(\xi)\,\left|\overline{U}_{\rm d}(q_\|,T,\xi)\right|^2\ ,
\label{eqn-7}
\end{eqnarray}
where $\phi_{{\bf q}_\|,{\bf p}_\|}$ is the angle between two wave vectors $\mbox{\boldmath$q$}_\|$ and $\mbox{\boldmath$p$}_\|$, and from Eq.\,\eqref{c-7} the defect interaction
$\left|\overline{U}_{\rm d}(q_\|,T,\xi)\right|^2$ with electrons is calculated as

\begin{eqnarray}
\nonumber
&&\left|\overline{U}_{\rm d}(q_\|,T,\xi)\right|^2\equiv\left.\int\limits_0^{\pi} d\phi\,\left|U_{\rm d}(\left|\mbox{\boldmath$q$}_\|-\mbox{\boldmath$p$}_\|\right|,T,\xi)\right|^2\right|_{q_\|=p_\|}\\
\nonumber
&=&\left(\frac{a_0}{a_{\rm d}}\right)^2\left[\frac{Z_{\rm eff}(T,E^*)\,e^2}{2\epsilon_0\epsilon_{\rm s}}\right]^2\int\limits_0^{\pi} d\phi\,\frac{\texttt{e}^{-\Delta_0^2(q_\|,\phi)\Lambda_\|^2/2}}{[\Delta_0(q_\|,\phi)+q_{\rm s}]^2}\\
&\times&\left[\int\limits_{0}^{\infty} dz\int\limits_{-\infty}^\infty dz'\,\left|{\cal F}_{0}(z)\right|^2\left(\texttt{e}^{-\Delta_0(q_\|,\phi)|z-z'|}+\texttt{e}^{-\Delta_0(q_\|,\phi)|z+z'|}\right){\cal Q}_1(\Delta_0(q_\|,\phi),z'-\xi)\right]^2\ ,\ \ \ \
\label{eqn-8}
\end{eqnarray}
${\cal L}_0=a+2\ell_0$ is the unit-cell size of a single well, $a$ is the inter-well separation,
$Z_{\rm eff}(T,E^*)$ is the trapped charge number by defects to be determined by Eq.\,\eqref{c-add},
$\Delta_0(q_\|,\phi)=q_\|\,|\sin(\phi/2)|$,
$\Lambda_\|$ is the correlation length for randomly-distributed defects,
the partial form factor ${\cal Q}_1(q_\|,\xi)$ is presented in Eq.\,\eqref{c-8},
and $\rho_{\rm d}(\xi)$ stands for a one-dimensional density distribution of defects to be determined by defect generation and diffusion\,\cite{feigao}.
Here, the prefactor $(a_0/a_{\rm d})^2$ in Eq.\,\eqref{eqn-8} comes from the fact that we are considering a spherical void with a larger radius $a_{\rm d}$ for a void of many 
point vacancies with a point-vacancy smaller radius $a_0\ll a_{\rm d}$,\,\cite{ajss} which leads to a lot of occupied lower energy levels and then a very large value for $Z_{\rm eff}(T,E^*)$.
\medskip

The lowest-order approximate solution of Eq.\,(\ref{eqn-7}) can be obtained simply by replacing $\Gamma(p_\|,\omega)$ with $1$ on the right-hand side of this equation. Therefore, the leading correction to
$\Gamma(q_\|,\omega)\approx 1$ becomes proportional to the total number of defects or integral of $\rho_{\rm d}(\xi)\,|\overline{U}_{\rm d}(q_\|,T,\xi)|^2$ with respect to $\xi$.
In general, the exact solution of Eq.\,(\ref{eqn-7}) includes all the higher orders of the integral of $\rho_{\rm d}(\xi)\,|\overline{U}_{\rm d}(q_\|,T,\xi)|^2$ by going beyond the second-order Born approximation\,\cite{huang}.
\medskip

Based on the calculated $|\overline{U}_{\rm d}(q_\|,T,\xi)|^2$ in Eq.\,\eqref{eqn-8}, Equation\ (\ref{eqn-7}) can be applied 
to compute the dynamical defect-vertex correction $\Gamma(q_\|,\omega)$ with respect to unity
within the LA. In order to simulate the physical distribution of defects,\,\cite{feigao} we assume in this paper a regional form\,\cite{feigao}, i.e.,
$\displaystyle{\frac{\rho_{\rm d}(\xi)}{\kappa_0}=\rho_{\rm L}\,\Theta(-\xi-\ell_0)+\rho_{\rm R}\,\Theta(\xi-\ell_0)+\left[\rho_{\rm W}+\xi\left(\Delta\rho/2\ell_0\right)\right]\,\Theta(\ell_0-|\xi|)}$,
where $\Theta(x)$ is a unit-step function, $\kappa_0$ is a scaling factor, and $\rho_{\rm W},\,\rho_{\rm L},\,\rho_{\rm R},\,\Delta\rho$ are parameters for linear defect densities in different regions of a quantum well.
Equation\ (\ref{eqn-7}) can be solved exactly and we are able to find the scattering resonance from the peak in the density plot of

\begin{equation}
\Gamma(q_\|,\omega)=\left[1-\frac{2m^*}{\pi^2\hbar^2}\,\chi_1^{(0)}(q_\|,\omega)
\int\limits_{-{\cal L}_0/2}^{{\cal L}_0/2} d\xi\,\rho_{\rm d}(\xi)\,\left|\overline{U}_{\rm d}(q_\|,T,\xi)\right|^2\right]^{-1}
\label{eqn-9}
\end{equation}
within the ($\omega,q_\|$)-plane.
The strength of this scattering resonance decreases rapidly with increasing $q_\|$ due to reduced $|\overline{U}_{\rm d}(q_\|,T,\xi)|^2$ from the suppressed long-range scattering but increases with $\kappa_0$ nonlinearly.
\medskip

The calculated $\Gamma(q_\|,\omega)$ can be substituted into Eq.\,(\ref{eqn-4}) to obtain the intra-well RPA dielectric function modified by defects.
Graphically, the dispersion relations of intrasubband-plasmon modes appear as a peak
in the density plot for $\displaystyle{\frac{\left|{\rm Im}[\epsilon(q_\|,\omega)]\right|}{\left\{{\rm Re}[\epsilon(q_\|,\omega)]\right\}^2+\left\{{\rm Im}[\epsilon(q_\|,\omega)]\right\}^2}}$ in the ($\omega,q_\|$)-plane, where the dielectric function
$\epsilon(q_\|,\omega)$ has been given by Eq.\,(\ref{eqn-4}).
In the absence of defects, we will find one intrasubband-plasmon mode and one particle-hole continuum (i.e., ${\rm Im}[\chi_1^{(0)}(q_\|,\omega)]\neq 0$),
correspondingly, for a single occupied subband in a quantum well.
After introducing defects to the quantum well,
the dispersion of this plasmon mode will be modified for small $q_\|$ values due to an energy shift from the contribution of ${\rm Re}[\Gamma(q_\|,\omega)]$.
Such a modification to the plasmon dispersion can acquire a nonlinear $\kappa_0$ dependence for large $\kappa_0$ values (or higher defect densities).
\medskip

By writing $\displaystyle{\kappa(q_\|,\omega)=\frac{1}{\epsilon(q_\|,\omega)}}$ with this modified dielectric function, the resulting inverse dielectric function
can further be input into Eq.\,(\ref{eqn-1}) to compute related changes $\delta\chi_1(q_\|,\omega)$ in the screened polarization function of a single quantum well.
In this case, the defect-induced change $\delta{\rm Im}[\chi_1(q_\|,\omega)]$ will display a peak shifting towards lower
values of $\omega$ with increasing $\kappa_0$. Meanwhile, $\delta{\rm Im}[\chi_1(q_\|,\omega)]$ reduces significantly for a larger $q_\|$ value due to weakened scattering interaction.
It would also be interesting to find how the depolarization shift of a plasmon peak (i.e., ${\rm Im}[\chi_1(q_\|,\omega)]$ vs. ${\rm Im}[\chi_1^{(0)}(q_\|,\omega)]$ with $\Gamma(q_\|,\omega)\equiv 1$)
increases with $q_\|$, but it will not show up in $\delta{\rm Im}[\chi_1(q_\|,\omega)]$ for defect effects.
This pure plasmon depolarization shift towards a higher $\omega$ value results from a many-body screening effect and will be reduced by defect scattering.
\medskip

For a multi-quantum-well system, on the other hand,
its density-density response function $\chi_{j,j'}(q_\|,\omega)$ satisfies the following self-consistent dynamical equation\,\cite{feigao}

\begin{equation}
\chi_{j,j'}(q_\|,\omega)=\chi_1(q_\|,\omega)\,\delta_{j,j'}+\chi_1(q_\|,\omega)\sum_{j^{\prime\prime}(\neq j)=0}^N\,v_{\rm c}(ja,j^{\prime\prime}a\,\vert\,q_\|)\,\chi_{j^{\prime\prime},j'}(q_\|,\omega)\ ,
\label{eqn-10}
\end{equation}
where the integers $j=0,\,1,\,\cdots,\,N$ label $N+1$ different quantum wells, 
the summation over $j^{\prime\prime}$ excludes the intra-well term with $j^{\prime\prime}=j$, 
and the single-layer $\chi_1(q_\|,\omega)$ is determined from Eq.\,\eqref{eqn-1}.
Moreover, the {\em inter-well} Coulomb coupling $v_{\rm c}(z,z'\,\vert\,q_\|)$
between two electrons in Eq.\,\eqref{eqn-10}, including the image potentials, is calculated as\,\cite{book}

\begin{eqnarray}
\nonumber
v_{\rm c}(z,z'\,\vert\,q_\|)&=&\frac{\beta_0(q_\|)\,e^2}{2\epsilon_0\epsilon_{\rm s}(q_\|+q_{\rm s})}\left[\texttt{e}^{-q_\||z-z'|}+\alpha_0^2\,\texttt{e}^{-2q_\|{\cal L}_{\rm tot}}\,\texttt{e}^{q_\||z-z'|}\right.\\
&+&\left.\alpha_0\,\texttt{e}^{-q_\||z+z'|}+\alpha_0\,\texttt{e}^{-2q_\|{\cal L}_{\rm tot}}\,e^{q_\||z+z'|}\right]\ ,
\label{eqn-11}
\end{eqnarray}
where ${\cal L}_{\rm tot}=Na$, $\displaystyle{\alpha_0=\frac{\epsilon_{\rm s}-1}{\epsilon_{\rm s}+1}}$, and
$\beta_0(q_\|)=\left[1-\alpha_0^2\exp(-2q_\|{\cal L}_{\rm tot})\right]^{-1}$. By using the  calculated density-density response function $\chi_{j,j'}(q_\|,\omega)$ in Eq.\,\eqref{eqn-10}, we are able to find the optical conductivity.\,\cite{new4,new5}

\section{Magneto-Transport in a Quantum Well}
\label{sec-3}

For an $n$-doped semiconductor quantum well,
we begin with the standard semi-classical Boltzmann transport equation (BTE) for conduction electrons within the lowest subband $\varepsilon_{k_\|}$ in the electric-quantum limit.
For this case, under the condition $\Delta E_{n,k_\|}\gg k_BT$, the non-equilibrium electron distribution function $f({\mbox{\boldmath$r$}}_\|,{\mbox{\boldmath$k$}}_\|;t)$ in position-momentum space satisfies\,\cite{jo1,jo2}

\begin{eqnarray}
\nonumber
&&\frac{\partial f({\mbox{\boldmath$r$}}_\|,{\mbox{\boldmath$k$}}_\|;t)}{\partial t}+\Big<\frac{d{\mbox{\boldmath$r$}}_\|(t)}{dt}\Big>_{\rm av}\cdot\mbox{\boldmath$\nabla$}_{{\bf r}_\|}f({\mbox{\boldmath$r$}}_\|,{\mbox{\boldmath$k$}}_\|;t)+\Big<\frac{d{\mbox{\boldmath$k$}_\|}(t)}{dt}\Big>_{\rm av}\cdot\mbox{\boldmath$\nabla$}_{{\bf k}_\|}f({\mbox{\boldmath$r$}}_\|,{\mbox{\boldmath$k$}}_\|;t)\\
\nonumber
&=&\left.\frac{\partial f({\mbox{\boldmath$r$}}_\|,{\mbox{\boldmath$k$}}_\|;t)}{\partial t}\right|_{\rm coll}
-2\,\sum_{n=1}^{\infty}\,n^2{\cal T}_n(k_\|,T)\left\{f({\mbox{\boldmath$r$}}_\|,{\mbox{\boldmath$k$}}_\|;t)\left(1-{\cal P}_n\right)\right.\\
\nonumber
&-&\left.\left[1-f({\mbox{\boldmath$r$}}_\|,{\mbox{\boldmath$k$}}_\|;t)\right]{\cal P}_n\,
\texttt{e}^{-\Delta E_{n,k_\|}/k_BT}\right\}\\
&\approx&\left.\frac{\partial f({\mbox{\boldmath$r$}}_\|,{\mbox{\boldmath$k$}}_\|;t)}{\partial t}\right|_{\rm coll}
-{\cal R}_{\rm c}(k_\|,T,E^*)\,f({\mbox{\boldmath$r$}}_\|,{\mbox{\boldmath$k$}}_\|;t)\ ,
\label{eqn-12}
\end{eqnarray}
where ${\mbox{\boldmath$r$}}_\|$ is a two-dimensional position vector, ${\mbox{\boldmath$k$}}_\|$ is a two-dimensional wave vector,
the first term at the right-hand side of the above equation represents the collision contributions of electrons with defects, phonons and other electrons,
$\displaystyle{\Delta E_{n,k_\|}=\frac{\hbar\omega_0}{2}+\varepsilon_{k_\|}+|E_n|}$,
$E_n$ is the defect $n$th energy level,
${\cal P}_n$ the defect $n$th energy-level occupation,
and the last step only holds for $k_BT\ll\Delta E_{n,k_\|}$. Moreover,
${\cal T}_n(k_\|,T)$ is the transition rate which can be found from Eqs.\,\eqref{eqn-35} and \eqref{new-2}.
Here, the conduction-electron {\em capture rate} ${\cal R}_{\rm c}(k_\|,T,E^*)$ introduced in the last term of Eq.\,\eqref{eqn-12} is calculated according to

\begin{equation}
{\cal R}_{\rm c}(k_\|,T,E^*)=2\,\sum_{n=1}^\infty\,n^2\,{\cal T}_n(k_\|,T)\left[1-f_0(E_n-E^*)\right]\ .
\label{new-36}
\end{equation}
Furthermore, for subband electrons,
we can define, in a semi-classical way, their group velocity through $\displaystyle{\Big<\frac{d{\mbox{\boldmath$r$}}_\|(t)}{dt}\Big>_{\rm av}
\equiv{\mbox{\boldmath$v$}}({\mbox{\boldmath$k$}}_\|)=\frac{1}{\hbar}\,\mbox{\boldmath$\nabla$}_{{\bf k}_\|}\varepsilon_{k_\|}}$.
Finally, we introduce the semi-classical Newton-type force equation for the wave vector of subband electrons, yielding

\begin{equation}
\hbar\,\Big<\frac{d{\mbox{\boldmath$k$}_\|}(t)}{dt}\Big>_{\rm av}={\mbox{\boldmath$F$}}({\mbox{\boldmath$k$}}_\|,t)\equiv-e\left[{\mbox{\boldmath$E$}}(t)+{\mbox{\boldmath$v$}}({\mbox{\boldmath$k$}}_\|)\times {\mbox{\boldmath$B$}}(t)\right]\ ,
\label{eqn-13}
\end{equation}
where ${\mbox{\boldmath$E$}}(t)$ and ${\mbox{\boldmath$B$}}(t)$ are the external time-dependent electric and magnetic fields, respectively, while ${\mbox{\boldmath$F$}}({\mbox{\boldmath$k$}}_\|,t)$
represents a microscopic electro-magnetic force acting on a single electron in the ${\mbox{\boldmath$k$}}_\|$ state.
\medskip

Based on Eq.\,(\ref{eqn-12}), one acquires the zeroth-order moment of BTE by summing over all the ${\mbox{\boldmath$k$}}_\|$ states on both sides of this equation.
This directly leads to the generalized electron number conservation law, i.e.,

\begin{equation}
\frac{\partial\rho_{\rm e}(\mbox{\boldmath$r$}_\|,\,t)}{\partial t}+\mbox{\boldmath$\nabla$}_{{\bf r}_\|}\cdot{\mbox{\boldmath$J$}}_{\rm e}(\mbox{\boldmath$k$}_\|,\,t)\\
=-\overline{\cal R}_{\rm c}(T,u_0,E^*)\,\rho_{\rm e}(\mbox{\boldmath$r$}_\|,t)\ ,
\label{eqn-14}
\end{equation}
where the electron areal density $\rho_{\rm e}({\mbox{\boldmath$r$}}_\|,t)$, the sheet current density ${\mbox{\boldmath$J$}}_{\rm e}({\mbox{\boldmath$r$}}_\|,t)$
and the captured-electron density $\overline{\cal R}_{\rm c}(T,u_0,E^*)\,\rho_{\rm e}({\mbox{\boldmath$r$}}_\|,t)$ are defined respectively by

\begin{eqnarray}
\rho_{\rm e}(\mbox{\boldmath$r$}_\|\,t)&=&\frac{2}{{\cal A}}\sum_{{\bf k}_\|}\,f({\mbox{\boldmath$r$}}_\|,{\mbox{\boldmath$k$}}_\|;t)\ ,\\
\label{eqn-15}
{\mbox{\boldmath$J$}}_{\rm e}({\mbox{\boldmath$r$}}_\|,t)&=&\frac{2}{{\cal A}}\sum_{{\bf k}_\|}\,{\mbox{\boldmath$v$}}({\mbox{\boldmath$k$}}_\|)\,f({\mbox{\boldmath$r$}}_\|,\,{\mbox{\boldmath$k$}}_\|;t)\ ,\\
\label{eqn-16}
\overline{\cal R}_{\rm c}(T,u_0,E^*)\,\rho_{\rm e}(\mbox{\boldmath$r$}_\|,t)&=&\frac{2}{{\cal A}}\sum_{{\bf k}_\|}\,{\cal R}_{\rm c}(k_\|,T,E^*)\,f({\mbox{\boldmath$r$}}_\|,{\mbox{\boldmath$k$}}_\|;t)\ ,\\
\label{new-23}
\overline{\cal R}_{\rm c}(T,u_0,E^*)&=&\frac{2}{\rho_0(u_0){\cal A}}\,\sum_{{\bf k}_\|}\,{\cal R}_{\rm c}(k_\|,T,E^*)f_0(\varepsilon_{k_\|}-u_0)\ ,
\label{new-24}
\end{eqnarray}
$\rho_0(u_0)=N_{\rm e}/{\cal A}$, $N_{\rm e}$ is the total number of subband electrons in the system (excluding the captured electrons by defects),
${\cal A}$ is the quantum-well cross-sectional area, the spin-degeneracy of electrons is included, and $\overline{\cal R}_{\rm c}(T,u_0,E^*)$ represents the statistical average of the capture rate.
\medskip

For the first-order moment of BTE, however, we have to employ the so-called Fermi kinetics\,\cite{jo1,jo2}. Therefore, we first introduce the relaxation-time approximation (RTA) for subband-electron collisions,
which conserves the total subband-electron number and is given by

\begin{equation}
\left.\frac{\partial f({\mbox{\boldmath$r$}}_\|,{\mbox{\boldmath$k$}}_\|;t)}{\partial t}\right|_{\rm coll}=-\,\frac{f({\mbox{\boldmath$r$}}_\|,{\mbox{\boldmath$k$}}_\|;t)-f_0(\varepsilon_{k_\|}-u_0)}{\tau_{\rm e}(k_\|)}\ ,
\label{eqn-17}
\end{equation}
where $\tau_{\rm e}(k_\|)$ describes the total energy-relaxation time for electrons in the ${\mbox{\boldmath$k$}}_\|$ state, including electron-electron, electron-phonon and electron-defect interactions.
After a time longer than the defect relaxation and capture times (see Sec.\,\ref{sec-5}), the chemical potential $u_0(T)$ of the subband-electron system can be determined self-consistently from

\begin{equation}
N_{\rm e}=2\,\sum_{{\bf k}_\|}\,f_0(\varepsilon_{k_\|}-u_0)=\int d^2{\mbox{\boldmath$r$}}_\|\,\rho({\mbox{\boldmath$r$}}_\|,t)
=\frac{2}{{\cal A}}\sum_{{\bf k}_\|}\,
\int d^2{\mbox{\boldmath$r$}}_\|\,f({\mbox{\boldmath$r$}}_\|,{\mbox{\boldmath$k$}}_\|;t)\ .
\label{eqn-18}
\end{equation}
Finally, by applying the RTA to the BTE in Eq.\,(\ref{eqn-12}), we find

\begin{eqnarray}
\nonumber
&&f({\mbox{\boldmath$r$}}_\|,{\mbox{\boldmath$k$}}_\|;t)+\tau_0(T,u_0)\,\frac{\partial f({\mbox{\boldmath$r$}}_\|,{\mbox{\boldmath$k$}}_\|;t)}{\partial t}
+\tau_0(T,u_0)\,{\cal R}_{\rm c}(k_\|,T,E^*)\,f({\mbox{\boldmath$r$}}_\|,{\mbox{\boldmath$k$}}_\|;t)\\
\nonumber
&\approx& f_0(\varepsilon_{k_\|}-u_0)
-\frac{\tau_0(T,u_0)}{\hbar}\,{\mbox{\boldmath$F$}}({\mbox{\boldmath$k$}}_\|,t)\cdot\mbox{\boldmath$\nabla$}_{{\bf k}_\|}f_0(\varepsilon_{k_\|}-u_0)
-\tau_0(T,u_0)\,\mbox{\boldmath$\nabla$}_{{\bf r}_\|}\cdot\left\{{\mbox{\boldmath$v$}}({\mbox{\boldmath$k$}}_\|)\,f_0(\varepsilon_{k_\|}-u_0)\right\}\\
&=&f_0(\varepsilon_{k_\|}-u_0)
-\frac{\tau_0(T,u_0)}{\hbar}\,{\mbox{\boldmath$F$}}({\mbox{\boldmath$k$}}_\|,t)\cdot\mbox{\boldmath$\nabla$}_{{\bf k}_\|}f_0(\varepsilon_{k_\|}-u_0)\ ,
\label{eqn-19}
\end{eqnarray}
where we have assumed that both $T$ and $u_0$ are spatially uniform throughout the system.
Additionally, $\tau_0(T,u_0)$ introduced in the above equation represents the statistically average of the energy-relaxation time, and is given by

\begin{equation}
\frac{1}{\tau_0(T,u_0)}=\frac{2}{N_{\rm e}}\sum_{{\bf k}_\|}\,\frac{f_0(\varepsilon_{k_\|}-u_0)}{\tau_{\rm e}(k_\|)}=
\frac{2}{N_{\rm e}}\sum_{{\bf k}_\|}\,\frac{1}{\tau_{\rm d}(k_\|)}\,f_0(\varepsilon_{k_\|}-u_0)\ ,
\label{eqn-20}
\end{equation}
where we have neglected the intrinsic contributions from Coulomb (for low electron densities) and phonons (for low temperatures) scattering in doped semiconductors,
and $\tau_{\rm d}(k_\|)$ is attributed to the extrinsic scattering between electrons and defects.
The explicit expression for $\tau_{\rm d}(k_\|)$ can be found from Eq.\,\eqref{a-1}.
\medskip

Now, let us further introduce an inverse momentum-relaxation time tensor $\tensor{\mbox{\boldmath$\tau$}}_p^{-1}$ in connection with Eq.\,(\ref{eqn-13}).
For this purpose, we would like to employ first the so-called force-balance equation\,\cite{jo2,r3} for a time-dependent
macroscopic drift velocity $\mbox{\boldmath$v$}_d(t)$ under external electric and magnetic fields, ${\mbox{\boldmath$E$}}(t)$ and ${\mbox{\boldmath$B$}}(t)$, yielding

\begin{eqnarray}
\nonumber
\frac{d\mbox{\boldmath$v$}_d(t)}{dt}&=&-\tensor{\mbox{\boldmath$\tau$}}_p^{-1}\cdot\mbox{\boldmath$v$}_d(t)+\tensor{\mbox{\boldmath$\cal M$}}^{-1}\cdot{\mbox{\boldmath$F$}}_{\rm e}(t)\\
&=&-\tensor{\mbox{\boldmath$\tau$}_p}^{-1}\cdot\mbox{\boldmath$v$}_d(t)
-e\tensor{\mbox{\boldmath${\cal M}$}}^{-1}\cdot\left[{\mbox{\boldmath$E$}}(t)+\mbox{\boldmath$v$}_d(t)\times{\mbox{\boldmath$B$}}(t)\right]=0\ ,
\label{eqn-21}
\end{eqnarray}
where ${\mbox{\boldmath$F$}}_{\rm e}(t)=-e\left[{\mbox{\boldmath$E$}}(t)+\mbox{\boldmath$v$}_d(t)\times{\mbox{\boldmath$B$}}(t)\right]$ is the macroscopic electro-magnetic force, while 
the statistically-averaged inverse effective-mass tensor $\tensor{\mbox{\boldmath${{\cal M}}$}}^{-1}$ for {\em isotropic} subband is

\begin{equation}
{\cal M}_{ij}^{-1}=\frac{2}{N_{\rm e}\hbar^2}\,\sum_{{\bf k}_\|}\left(\frac{\partial^2\varepsilon_{k_\|}}{\partial k_i\partial k_j}\right)f_0(\varepsilon_{k_\|}-u_0)
=\delta_{ij}\,\frac{2}{N_{\rm e}\hbar^2}\,\sum_{{\bf k}_\|}\left(\frac{\partial^2\varepsilon_{k_\|}}{\partial^2k_\|}\right)f_0(\varepsilon_{k_\|}-u_0)\ .
\label{eqn-22}
\end{equation}
The explicit form of $\tensor{\mbox{\boldmath$\tau$}}_p^{-1}$ in our system will be presented in Sec.\,\ref{sec-4} next.
As shown in Appendix\ \ref{app-2}, the solution of Eq.\,(\ref{eqn-21}) can be formally written as

\begin{equation}
\mbox{\boldmath$v$}_d(t)=\tensor{\mbox{\boldmath${\mu}$}}[{\mbox{\boldmath$B$}}(t)]\cdot{\mbox{\boldmath$E$}}(t)\ ,
\label{eqn-23}
\end{equation}
where $\tensor{\mbox{\boldmath${\mu}$}}[{\mbox{\boldmath$B$}}(t)]$ is the so-called mobility tensor\,\cite{jo2} for describing magneto-transport of subband electrons,
which also depends on $\tensor{\mbox{\boldmath$\tau$}}_p^{-1}$ and $\tensor{\mbox{\boldmath${\cal M}$}}^{-1}$ in addition to ${\mbox{\boldmath$B$}}(t)$.
The details for calculating $\tensor{\mbox{\boldmath${\mu}$}}[{\mbox{\boldmath$B$}}(t)]$ can be found in Appendix \ref{app-2}.
Using Eqs.\,(\ref{eqn-21}) and (\ref{eqn-23}), we are able to simply rewrite the macroscopic electro-magnetic force as
${\mbox{\boldmath$F$}}_{\rm e}(t)=\left(\tensor{\mbox{\boldmath${\cal M}$}}\otimes\tensor{\mbox{\boldmath$\tau$}_p}^{-1}\right)\cdot\left[\tensor{\mbox{\boldmath${\mu}$}}[{\mbox{\boldmath$B$}}(t)]\cdot{\mbox{\boldmath$E$}}(t)\right]$, where
$\tensor{\mbox{\boldmath${\cal M}$}}$ represents the inverse of the tensor $\tensor{\mbox{\boldmath${\cal M}$}}^{-1}$.
\medskip

In a similar way with deriving Eq.\,\eqref{eqn-14}, multiplying both sides of Eq.\,(\ref{eqn-19}) by the group velocity ${\mbox{\boldmath$v$}}({\mbox{\boldmath$k$}}_\|)$ and summing over all the ${\mbox{\boldmath$k$}}_\|$ states afterwards, we arrive at

\begin{eqnarray}
\nonumber
&&\left[1+s_{\rm c}(T,u_0,E^*)\right]\mbox{\boldmath$J$}_0(t)+\tau_0(T,u_0)\,\frac{d\mbox{\boldmath$J$}_0(t)}{dt}\\
\nonumber
&=&-\tau_0(T,u_0)\,\frac{2}{{\cal A}}\,\sum_{{\bf k}_\|}\,{\mbox{\boldmath$v$}}({\mbox{\boldmath$k$}}_\|)
\left[{\mbox{\boldmath$F$}}_{\rm e}(t)\cdot{\mbox{\boldmath$v$}}({\mbox{\boldmath$k$}}_\|)\right]\frac{\partial f_0(\varepsilon_{k_\|}-u_0)}{\partial\varepsilon_{k_\|}}\\
&=&\tau_0(T,u_0)\frac{2}{{\cal A}}\,\sum_{{\bf k}_\|}\,{\mbox{\boldmath$v$}}({\mbox{\boldmath$k$}}_\|)\left\{\left(\tensor{\mbox{\boldmath${\cal M}$}}\otimes\tensor{\mbox{\boldmath${\tau}$}_p}^{-1}\right)\cdot\left[\tensor{\mbox{\boldmath${\mu}$}}[{\mbox{\boldmath$B$}}(t)]\cdot{\mbox{\boldmath$E$}}(t)\right]\right\}\cdot{\mbox{\boldmath$v$}}({\mbox{\boldmath$k$}}_\|)\,\left[-\frac{\partial f_0(\varepsilon_{k_\|}-u_0)}{\partial\varepsilon_{k_\|}}\right]\ ,\ \ \ \ \ \
\label{eqn-24}
\end{eqnarray}
where the first term on the left-hand side of this equation includes the captured sheet current density $s_{\rm c}(T,u_0,E^*)\mbox{\boldmath$J$}_0(t)$,
$s_{\rm c}(T,u_0,E^*)=\overline{\cal R}_{\rm c}(T,u_0,E^*)\tau_0(T,u_0)$ is the current-suppression factor
(historically also called the capture probability if $s_{\rm c}(T,u_0,E^*)\ll1$), and
the second term on the left-hand side of the equation represents the non-adiabatic correction to $\mbox{\boldmath$J$}_0(t)$,
i.e., $\displaystyle{\frac{d{\mbox{\boldmath$J$}}_0(t)}{dt}=-i\omega\tilde{\mbox{\boldmath$J$}}_0}$ if ${\mbox{\boldmath$E$}}(t)=\tilde{\mbox{\boldmath$E$}}_0\texttt{e}^{-i\omega t}$ is assumed.
From Eq.\,(\ref{eqn-24}) we know the number current density $\mbox{\boldmath$J$}_0$ is independent of $\mbox{\boldmath$r$}_\|$
within the RTE. Consequently, from Eq.\,(\ref{eqn-14}) we find that the number areal density $\rho_e$
remains to be a constant $\rho_0(u_0)$ after a time longer than the defect energy-relaxation and capture times, and is determined by

\begin{equation}
\rho_0(u_0)=n_{\rm dop}-\rho_{\rm cap}(T,u_0,E^*)=\frac{2}{{\cal A}}\sum_{{\bf k}_\|}\,f_0(\varepsilon_{k_\|}-u_0)\equiv\frac{N_{\rm e}}{{\cal A}}\ ,
\label{eqn-25}
\end{equation}
which should self-consistently determine the chemical potential $u_0(T)$ of the system at any given $T$. In Eq.\,\eqref{eqn-25}, $n_{\rm dop}$ is the given extrinsic areal doping density,
$E^*$ the Fermi energy for bound electrons inside thermal-equilibrium defects,
and $\rho_{\rm cap}(T,u_0,E^*)$ represents the captured areal electron density as calculated by Eq.\,\eqref{new-5}.
If the external fields are static, i.e., ${\mbox{\boldmath$E$}}_0$ and ${\mbox{\boldmath$B$}}_0$, we get the sheet charge current density
$\mbox{\boldmath$j$}_0\equiv -e\tilde{\mbox{\boldmath$J$}}_0$ from Eq.\,(\ref{eqn-24}), i.e.,

\begin{eqnarray}
\nonumber
\mbox{\boldmath$j$}_0&=&-\frac{e\tau_0(T,u_0)}{1+s_{\rm c}(T,u_0,E^*)}\,\frac{2}{{\cal A}}\sum_{{\bf k}_\|}\,\mbox{\boldmath$v$}({\mbox{\boldmath$k$}}_\|)\\
&\times&
\left\{\left(\tensor{\mbox{\boldmath${\cal M}$}}\otimes\tensor{\mbox{\boldmath${\tau}$}}_p^{-1}\right)\cdot\left[\tensor{\mbox{\boldmath${\mu}$}}({\mbox{\boldmath$B$}}_0)\cdot{\mbox{\boldmath$E$}}_0)\right]\right\}
\cdot\mbox{\boldmath$v$}({\mbox{\boldmath$k$}}_\|)\,
\left[-\frac{\partial f_0(\varepsilon_{k_\|}-u_0)}{\partial\varepsilon_{k_\|}}\right]\ .
\label{eqn-26}
\end{eqnarray}
In this case, the elements of a conductivity tensor $\tensor{\mbox{\boldmath$\sigma$}}(\mbox{\boldmath$B$}_0)$ can be obtained through the relation
$\displaystyle{\sigma_{ij}(\mbox{\boldmath$B$}_0)=\frac{\mbox{\boldmath$j$}_0\cdot\hat{\mbox{\boldmath$e$}}_i}{\mbox{\boldmath$E$}_0\cdot\hat{\mbox{\boldmath$e$}_j}}}$, where $i,\,j=x,\,y$ and $\hat{\mbox{\boldmath$e$}}_x,\,\hat{\mbox{\boldmath$e$}}_y$
are two unit vectors
in two-dimensional position space.
From Eq.\,(\ref{eqn-26}), we further know that the conductivity tensor depends not only on the mobility tensor, but also on how electrons are thermally-distributed within a subband.
\medskip

As a special case, we consider an isotropic parabolic subband written as $\displaystyle{\varepsilon_{k_\|}=\frac{\hbar^2k_\|^2}{2m^\ast}}$,
we find from Eq.\,(\ref{eqn-22}) that $\displaystyle{{\cal M}_{ij}^{-1}=\frac{1}{m^\ast}\,\delta_{ij}}$, ${\cal M}_{ij}=m^\ast\,\delta_{ij}$,
and $\displaystyle{(\tensor{\tau_{p}}^{-1})_{ij}=\frac{1}{\tau_p}\,\delta_{ij}}$. In this case, from Eq.\,\eqref{b-12} we get the mobility tensor

\begin{equation}
\tensor{\mbox{\boldmath${\mu}$}}({\mbox{\boldmath$B$}}_0)=-\frac{\mu_0}{1+\mu_0^2B_z^2}\,
\left[\begin{array}{cc}
1 & -\mu_0B_z\\
\\
\mu_0B_z & 1
\end{array}\right]\ ,
\label{eqn-27}
\end{equation}
where $\displaystyle{\mu_0=\frac{e\tau_p}{m^\ast}}$ for ${\mbox{\boldmath$E$}}_0=\{E_x,\,E_y,\,0\}$ and ${\mbox{\boldmath$B$}}_0=\{0,\,0,\,B_z\}$. If we further assume ${\mbox{\boldmath$B$}}_0=0$, Eq.\,(\ref{eqn-27}) simply reduces to
$\mu_{ij}=-\mu_0\,\delta_{ij}$. In this case, from Eq.\,(\ref{eqn-26}) we recover the well-known Ohm's law
$\displaystyle{{\mbox{\boldmath$j$}}_0=\frac{\rho_0e^2\tau_0}{m^\ast}\,{\mbox{\boldmath$E$}}_0}$ after setting $s_{\rm c}(T,u_0,E^*)=0$, which implies $\displaystyle{\sigma_{ij}=\frac{\rho_0e^2\tau_0}{m^\ast}\,\delta_{ij}}$.
\medskip

\section{Momentum Dissipation from Defect Scattering}
\label{sec-4}

The inverse momentum-relaxation-time tensor $\tensor{\mbox{\boldmath${\tau}$}}_p^{-1}$, first introduced in Eq.\,(\ref{eqn-21}), comes from a statistically-averaged resistive forces ${\mbox{\boldmath${\cal F}$}}_p$
due to elastic scattering of electrons with randomly-distributed defects in a quantum well.\,\cite{jo2}
\medskip

For field-driven electrons moving with a drift velocity $\mbox{\boldmath$v$}_d$, the total resistive force ${\mbox{\boldmath${\cal F}$}}_p$, which can be evaluated from electron momentum dissipation due to
defect elastic scattering, is calculated in the quasi-elastic limit as\,\cite{jo2}

\begin{equation}
{\mbox{\boldmath${\cal F}$}}_p=-\frac{2\pi N_{\rm d}}{\hbar {\cal A}^2}\,\sum_{{\bf k}_\|,{\bf q}_\|}\,\hbar{\mbox{\boldmath$q$}}_\|\left(\hbar{\mbox{\boldmath$q$}}_\|\cdot\mbox{\boldmath$v$}_d\right)\,\left|U_0(q_\|)\right|^2\,
\left(-\frac{\partial f_{{\bf k}_\|}}{\partial\varepsilon_{{\bf k}_\|}}\right)\,
\delta(\varepsilon_{{\bf k}_\|+{\bf q}_\|}-\varepsilon_{{\bf k}_\|}+\hbar{\mbox{\boldmath$q$}}_\|\cdot\mbox{\boldmath$v$}_d)\ ,
\label{eqn-28}
\end{equation}
where $N_{\rm d}$ is the total number of defects, and 

\begin{eqnarray}
\nonumber
&&\left|U_0(q_\|)\right|^2=\left(\frac{a_0}{a_{\rm d}}\right)^2\left[\frac{Z_{\rm eff}(T,E^*)\,e^2}{2\epsilon_0\epsilon_{\rm s}}\right]^2\,
\frac{\texttt{e}^{-q^2_\|\Lambda^2_\|/2}}{(q_\|+q_{\rm s})^2}\\
&\times&
\int\limits_{-{\cal L}_0/2}^{{\cal L}_0/2} d\xi\,\rho_{\rm d}(\xi)
\left[\int\limits_{0}^{\infty} dz\int\limits_{-\infty}^\infty dz'\,\left|{\cal F}_{0}(z)\right|^2\left(\texttt{e}^{-q_\||z-z'|}+\texttt{e}^{-q_\||z+z'|}\right){\cal Q}_1(q_\|,z'-\xi)\right]^2\ .\ \ \ \ \ \
\label{eqn-29}
\end{eqnarray}
Here, for simplicity we have introduced the notation $f_{{\bf k}_\|}\equiv f_0(\varepsilon_{k_\|}-u_0)$ in Eq.\,\eqref{eqn-28}.
Using the relation $\displaystyle{\tensor{\mbox{\boldmath${\tau}$}}_{p}^{-1}\cdot\mbox{\boldmath$v$}_d=-\frac{2}{N_{\rm e}}\,\tensor{\mbox{\boldmath${\cal M}$}}^{-1}\cdot{\mbox{\boldmath${\cal F}$}}_p}$, as obtained
from the force-balance equation in Eq.\,\eqref{eqn-21},
we get from Eq.\,\eqref{eqn-28} the inverse momentum-relaxation-time tensor $\tensor{\mbox{\boldmath${\tau}$}}_p^{-1}$
in the quasi-elastic limit, yielding,

\begin{equation}
\tensor{\mbox{\boldmath${\tau}$}}_{p}^{-1}=\frac{4\pi\hbar\,\sigma_{\rm d}}{\rho_0(u_0){\cal A}^2}\,\sum_{{\bf k}_\|,{\bf q}_\|}\,\left|U_0(q_\|)\right|^2\,
\left(-\frac{\partial f_{{\bf k}_\|}}{\partial\varepsilon_{{\bf k}_\|}}\right)\,\delta(\varepsilon_{{\bf k}_\|+{\bf q}_\|}-\varepsilon_{{\bf k}_\|}+\hbar{\mbox{\boldmath$q$}}_\|\cdot\mbox{\boldmath$v$}_d)\,\left\{\tensor{\mbox{\boldmath${\cal M}$}}^{-1}\otimes\left[{\mbox{\boldmath$q$}}_\|\otimes{\mbox{\boldmath$q$}}_\|^T\right]\right\}\ ,
\label{eqn-30}
\end{equation}
where $\sigma_{\rm d}=N_{\rm d}/{\cal A}$ is the point-defect areal density, and the $2\times 2$ matrix is defined as

\[
\left[{\mbox{\boldmath$q$}}_\|\otimes{\mbox{\boldmath$q$}}_\|^T\right]\equiv\left[\begin{array}{cc}
q_x^2 & q_xq_y\\
\\
q_yq_x & q_y^2
\end{array}\right]\ .
\]
\medskip

Finally, by employing the expression in Eq.\,\eqref{eqn-2} and the density-density response function $\chi_{j,j'}(q_\|,\omega)$ determined from Eq.\,\eqref{eqn-10},
we can rewrite Eq.\,(\ref{eqn-28}) in a compact form, i.e.,

\begin{equation}
{\mbox{\boldmath${\cal F}$}}_p=\frac{N_{\rm d}}{{\cal A}}\sum_{{\bf q}_\|}\,{\mbox{\boldmath$q$}}_\|\left|U_0(q_\|,T)\right|^2
{\rm Im}[\tilde{\chi}(q_\|,\omega=-\mbox{\boldmath$q$}_\|\cdot\mbox{\boldmath$v$}_d)]\ ,
\label{eqn-31}
\end{equation}
where we have defined a total density-density response function $\tilde{\chi}(q_\|,\omega)$ for the whole multi-quantum-well system, given by $\displaystyle{\tilde{\chi}(q_\|,\omega)=\sum_{j=0}^{N}\,\chi_{j,j}(q_\|,\omega)}$.
By using Eq.\,(\ref{eqn-31}), the inverse momentum-relaxation-time tensor $\tensor{\mbox{\boldmath${\tau}$}}_{p}^{-1}$ presented in Eq.\,(\ref{eqn-30}) can also be rewritten in the quasi-elastic limit as

\begin{equation}
\tensor{\mbox{\boldmath${\tau}$}}_{p}^{-1}\approx \frac{2\sigma_{\rm d}}{\rho_0(u_0){\cal A}}\,\sum_{{\bf q}_\|}\,\left|U_0(q_\|)\right|^2\left\{\frac{\partial}{\partial\omega}\,{\rm Im}[\tilde{\chi}(q_\|,\omega)]\right\}_{\omega=0}
\left\{\tensor{\mbox{\boldmath${\cal M}$}}^{-1}\otimes\left[{\mbox{\boldmath$q$}}_\|\otimes{\mbox{\boldmath$q$}}_\|^T\right]\right\}\ .
\label{eqn-33}
\end{equation}
\medskip

Assuming $\displaystyle{\tensor{\mbox{\boldmath${\cal M}$}}^{-1}=\frac{1}{m^*}\,\tensor{\mbox{\boldmath${\cal I}$}}_0}$ with $\tensor{\mbox{\boldmath${\cal I}$}}_0$ as a unit matrix, we get from Eq.\,\eqref{eqn-33}
that $\displaystyle{\tensor{\mbox{\boldmath${\tau}$}}_{p}^{-1}=\frac{1}{\tau_p}\,\tensor{\mbox{\boldmath${\cal I}$}}_0}$ and

\begin{equation}
\frac{1}{\tau_p}=\frac{\sigma_{\rm d}}{m^*\rho_0(u_0){\cal A}}\,\sum_{{\bf q}_\|}\,q_\|^2\,\left|U_0(q_\|)\right|^2\left\{\frac{\partial}{\partial\omega}\,{\rm Im}[\tilde{\chi}(q_\|,\omega)]\right\}_{\omega=0}\ ,
\label{eqn-34}
\end{equation}
which relates to the so-called memory-function formalism\,\cite{huag1} in the static limit.
From the result in Eq.\,\eqref{eqn-34}, we are able to calculate the mobility tensor of this multi-quantum-well system using Eq.\,\eqref{eqn-27}, which includes the scattering contribution from defects in the system.

\section{Defect Capture and Charging Dynamics}
\label{sec-5}

\begin{figure}
\centering
\includegraphics[width=0.45\textwidth]{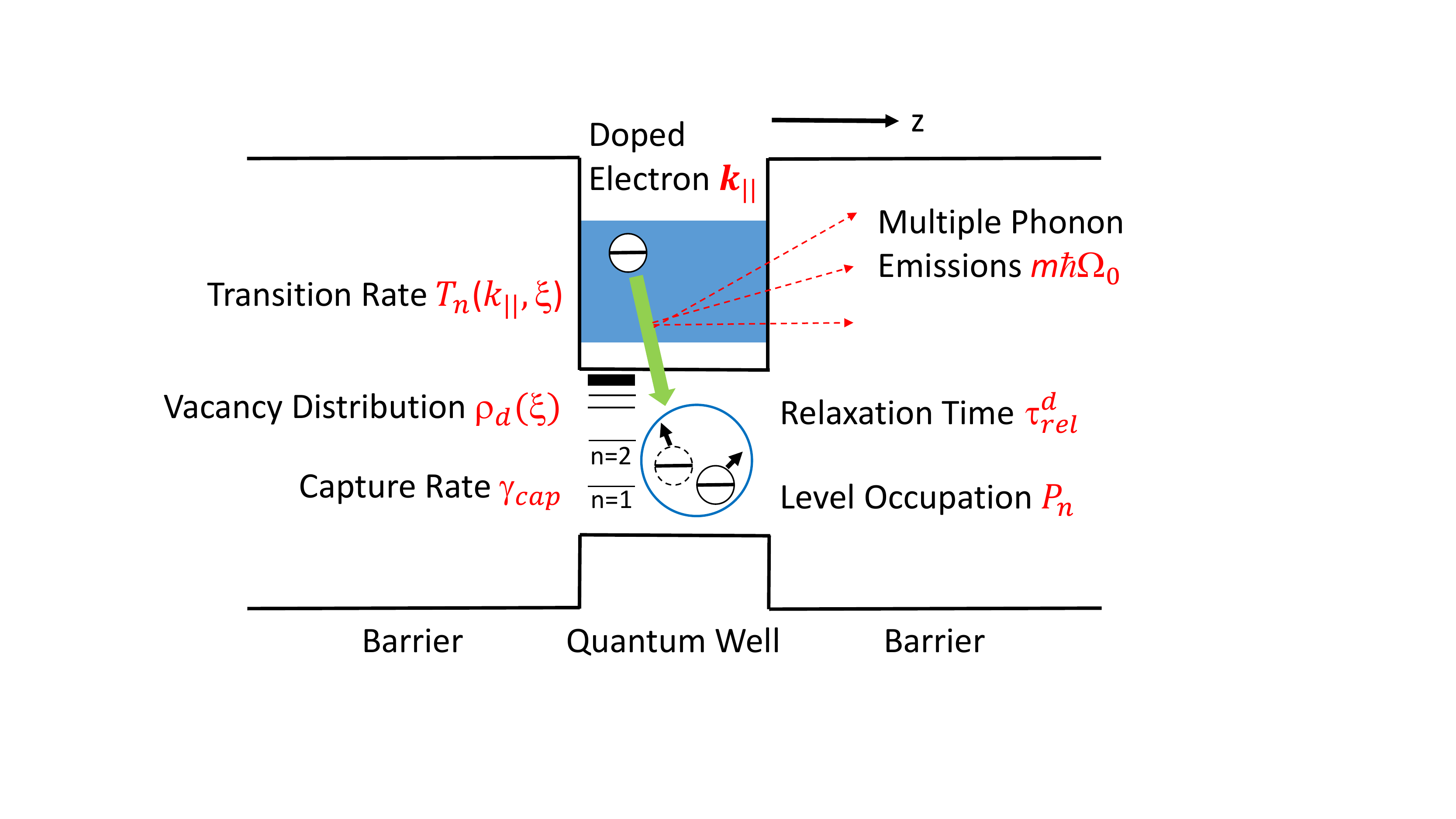}
\caption{Illustration for a spatial distribution $\rho_{\rm d}(\xi)$ of defects in a quantum-well structure,
where doped electrons in the Bloch $\mbox{\boldmath$k$}_\|$ state are captured by one defect sitting at $z=\xi$ with a rate $\gamma_{\rm cap}$ calculated from a phonon-mediated transition rate ${\cal T}_n(k_\|,\xi)$.
Meanwhile, multiple ($m$) optical phonons with total energy $m\hbar\Omega_0$ are emitted during this capturing process. This leads to a non-equilibrium level occupation ${\cal P}_n$ for the $n$th energy level
($n=1,\,2,\,\cdots$) of a defect,
accompanied by an energy-relaxation time $\tau^{\rm d}_{\rm rel}$.}
\label{fig1}
\end{figure}

As illustrated in Fig.\,\ref{fig1}, a higher-energy subband electron undergoes a capturing process by lower energy levels of a defect through a non-radiative decay
accompanied by multi-phonon emissions\,\cite{huang10}.
By including this multi-phonon emission mechanism\,\cite{huang6,ridley}, the transition rate between a Bloch $\mbox{\boldmath$k$}_\|$ state
and the $n$th energy level of a defect is calculated as\,\cite{multiphonon1,multiphonon2,hrfactor}

\begin{eqnarray}
\nonumber
{\cal T}_n(k_\|,T,\xi)&=&\frac{2\pi}{\hbar}\sum\limits_{\ell=0}^{n-1}\,\sum\limits_{m=-\ell}^{\ell}\,\exp\left[-(1+2N_0(\Omega_0))\,{\cal S}_0+\frac{\Delta E_{n,k_\|}}{2k_BT}\right]\,
\left(1-\frac{\Delta E_{n,k_\|}}{\hbar\Omega_0{\cal S}_0}\right)^2\frac{{\cal S}_0}{\hbar\Omega_0}\,\\
&\times&\sum_{\alpha=1}^{\infty}\,I_\alpha(\eta_0)\,
\delta\left(\alpha-\frac{\Delta E_{n,k_\|}}{\hbar\Omega_0}\right)\,
\left|\langle\psi_{n,\ell,m}(\mbox{\boldmath$r$}_\|,z-\xi)\vert
U_{\rm ep}(\mbox{\boldmath$r$}_\|,z)\vert\phi_{k_\|}(\mbox{\boldmath$r$}_\|,z)\rangle\right|^2\ ,\ \ \ \ \ \
\label{eqn-35}
\end{eqnarray}
where we have neglected the insignificant field-induced tunneling of trapped electrons by a defect,
$U_{\rm ep}(\mbox{\boldmath$r$}_\|,z)$ represents the potential energy due to electron-phonon coupling within a quantum well,
$\psi_{n,\ell,m}(\mbox{\boldmath$r$}_\|,z-\xi)$ is the wave function of a defect at the position $\xi$,
$\displaystyle{\phi_{k_\|}(\mbox{\boldmath$r$}_\|,z)={\cal F}_0(z)\,\frac{\exp(i\mbox{\boldmath$k$}_\|\cdot\mbox{\boldmath$r$}_\|)}{\sqrt{{\cal A}}}}$ is the wave function of
a Bloch electron in the quantum well, ${\cal S}_0$ the so-called Huang-Rhys factor\,\cite{bk1},
$\displaystyle{\Delta E_{n,k_\|}=\frac{\hbar\omega_0}{2}+\varepsilon_{k_\|}+|E_n|}$ the energy release by a captured electron,
$\hbar\Omega_0$ the longitudinal-optical-phonon energy, 
$N_0(\Omega_0)=\left[\exp(\hbar\Omega_0/k_{\rm B}T)-1\right]^{-1}$ the distribution function for thermal-equilibrium longitudinal-optical phonons, and $I_\alpha(\eta_0)$ is the modified Bessel function of order $\alpha$ with
$\eta_0=2{\cal S}_0\sqrt{N_0(\Omega_0)[N_0(\Omega_0)+1]}$.
\medskip

By utilizing the Fr\"ohlich Hamiltonian model,\,\cite{huang5} the electron-phonon coupling matrix element $\left|\langle\psi_{n,\ell,m}\vert U_{\rm ep}\vert\phi_{k_\|}\rangle\right|^2$ in Eq.\,(\ref{eqn-35}) is evaluated as

\begin{eqnarray}
\nonumber
&&\left|\langle\psi_{n,\ell,m}\vert U_{\rm ep}\vert\phi_{k_\|}\rangle\right|^2
=\frac{e^2}{2\epsilon_0\epsilon_{\rm d}}\left(\frac{1}{\epsilon_{\rm H}}-\frac{1}{\epsilon_{\rm L}}\right)\frac{\hbar\Omega_0}{2}\,\frac{1}{(2\pi)^2}
\int \frac{d^2\mbox{\boldmath$q$}_\|}{q_\|+q_s}
\int\limits_{-\infty}^\infty dz\int_{-\infty}^\infty dz'\,\texttt{e}^{-q_\||z-z'|}\,\\
&\times&
{\cal F}_0(z)\,{\cal B}^*_{n,\ell,m}(|\mbox{\boldmath$q$}_\|+\mbox{\boldmath$k$}_\||,z-\xi)\,
{\cal B}_{n,\ell,m}(|\mbox{\boldmath$q$}_\|+\mbox{\boldmath$k$}_\||,z'-\xi)\,{\cal F}_0(z')\ ,\ \ \ \
\label{eqn-36}
\end{eqnarray}
where $\epsilon_{\rm L}$ and $\epsilon_{\rm H}$ are the static and optic dielectric constants of the host semiconductor, while 
the Fourier-transformed partial form factor ${\cal B}_{n,\ell,m}(|\mbox{\boldmath$q$}_\|+\mbox{\boldmath$k$}_\||,z)$ of a defect is
	
\begin{eqnarray}
\nonumber
&&{\cal B}_{n,\ell,m}(|\mbox{\boldmath$q$}_\|+\mbox{\boldmath$k$}_\||,z)\equiv\int d^2{\bf r}_\|\,\psi_{n,\ell,m}(\mbox{\boldmath$r$}_\|,z)\,
\texttt{e}^{-i({\bf q}_\|+{\bf k}_\|)\cdot{\bf r}_\|}\\
&=&\sqrt{2\pi}\,\texttt{e}^{im\theta_{{\bf k}_\|+{\bf q}_\|}}(-i)^m\int\limits_0^\infty dr_\|\,r_\|J_m(|\mbox{\boldmath$q$}_\|+\mbox{\boldmath$k$}_\||r_\|)\,
R_{n,\ell}\left(\sqrt{r^2_\|+z^2}\right)\,\overline{Y}_{\ell,m}
\left(z/\sqrt{r_\|^2+z^2}\right)\ .
\label{eqn-37}
\end{eqnarray}
In Eq.\,\eqref{eqn-37},
$\psi_{n,\ell,m}(\mbox{\boldmath$r$}_\|,z)$ has been employed from Eq.\,\eqref{c-2}, $J_m(x)$ is the first-kind Bessel function of order $m$, and
$\displaystyle{\theta_{{\bf k}_\|+{\bf q}_\|}=\tan^{-1}\left(\frac{k_y+q_y}{k_x+q_x}\right)}$ is the angle between the wave vector $\mbox{\boldmath$k$}_\|+\mbox{\boldmath$q$}_\|=\{k_x+q_x,k_y+q_y\}$ and positive $x$-axis.
Here, the prefactor in front of the integral with respect to $r_\|$ in Eq.\,\eqref{eqn-37} only contributes a $2\pi$ factor in Eq.\,\eqref{eqn-36}.
\medskip

Based on the calculated transition rate ${\cal T}_n(k_\|,\xi)$ in Eq.\,\eqref{eqn-35} and
thermal-equilibrium energy-level occupation $f_0(E_n-E^*)$ in Eq.\,\eqref{c-9}, as well as using the defect-relaxation time approximation,
we arrive at the following dynamical equation for non-equilibrium occupation ${\cal P}_n$ of bound electrons on the $n$th energy level of a defect, i.e.,

\begin{eqnarray}
\nonumber
&&\left(1-{\cal P}_n\right)\frac{2}{\rho_0(u_0){\cal A}}\sum_{{\bf k}_\|}\,f_0(\varepsilon_{k_\|}-u_0)\,{\cal T}_n(k_\|,T)
-{\cal P}_n\,\frac{2}{\rho_0(u_0){\cal A}}\sum_{{\bf k}_\|}\,\left[1-f_0(\varepsilon_{k_\|}-u_0)\right]\\
\nonumber
&\times&\texttt{e}^{-\Delta E_{n,k_\|}/k_BT}\,{\cal T}_n(k_\|,T)+\left(1-{\cal P}_n\right)\sum_{n'>n}\,{\cal D}_{n,n'}(T)\,{\cal P}_{n'}
-{\cal P}_n\sum_{n>n'}\,{\cal D}_{n',n}(T)\left(1-{\cal P}_{n'}\right)\\
\nonumber
&+&\left(1-{\cal P}_n\right)\sum_{n'<n}\,{\cal D}_{n,n'}(T)\,{\cal P}_{n'}\,\texttt{e}^{-|E_{n}-E_{n'}|/k_BT}
-{\cal P}_n\sum_{n<n'}\,{\cal D}_{n',n}(T)\left(1-{\cal P}_{n'}\right)\texttt{e}^{-|E_{n}-E_{n'}|/k_BT}\\
&=&-\frac{{\cal P}_n-f_0(E_n-E^*)}{\tau^{\rm d}_n(T,u_0)}\ ,
\label{eqn-38}
\end{eqnarray}
where the non-radiative transition rate ${\cal D}_{n',n}(T)$ from the $n$th to the $n'$th energy level is

\begin{eqnarray}
\nonumber
&&{\cal D}_{n',n}(T)=\frac{2\pi}{\hbar}\,\exp\left\{-[1+2N_0(\Omega_0)]\,{\cal S}_0+\frac{|E_{n}-E_{n'}|}{2k_BT}\right\}\,
\left(1-\frac{|E_{n}-E_{n'}|}{\hbar\Omega_0{\cal S}_0}\right)^2\frac{{\cal S}_0}{\hbar\Omega_0}\,\\
&\times&\sum_{\alpha=1}^{\infty}\,I_\alpha(\eta_0)\,
\delta\left(\alpha-\frac{|E_{n}-E_{n'}|}{\hbar\Omega_0}\right)\,
\sum_{\ell,\ell',m,m'}\,\left|\langle\psi_{n',\ell',m'}(\mbox{\boldmath$r$}_\|,z)\vert
U_{\rm ep}(\mbox{\boldmath$r$}_\|,z)\vert\psi_{n,\ell,m}(\mbox{\boldmath$r$}_\|,z)\rangle\right|^2\ .\ \ \ \ \ \
\label{eqn-nonrad}
\end{eqnarray}
In addition, the defect-averaged transition rate ${\cal T}_n(k_\|,T,\xi)$ in Eq.\,\eqref{eqn-35} becomes

\begin{equation}
{\cal T}_n(k_\|,T)=\frac{1}{N_{\rm d}}\int\limits_{-{\cal L}_0/2}^{{\cal L}_0/2} d\xi\,\rho_{\rm d}(\xi)\,{\cal T}_n(k_\|,T,\xi)\ ,
\label{new-2}
\end{equation}
the $n$th-level {\em defect-relaxation time}, $\tau^{\rm d}_n(T,u_0,E^*)$, is given by

\begin{eqnarray}
\nonumber
&&\frac{1}{\tau^{\rm d}_n(T,u_0,E^*)}=\frac{2}{\rho_0(u_0){\cal A}}\sum_{{\bf k}_\|}\,{\cal T}_n(k_\|,T)\left\{f_0(\varepsilon_{k_\|}-u_0)+\left[1-f_0(\varepsilon_{k_\|}-u_0)\right]\texttt{e}^{-\Delta E_{n,k_\|}/k_BT}\right\}\\
\nonumber
&+&\sum_{n'>n}\,{\cal D}_{n,n'}(T)\,f_0(E_{n'}-E^*)+\sum_{n'<n}\,{\cal D}_{n,n'}(T)\,f_0(E_{n'}-E^*)\,\texttt{e}^{-|E_{n}-E_{n'}|/k_BT}\\
&+&\sum_{n>n'}\,{\cal D}_{n',n}(T)\left[1-f_0(E_{n'}-E^*)\right]+\sum_{n<n'}\,{\cal D}_{n',n}(T)\left[1-f_0(E_{n'}-E^*)\right]\texttt{e}^{-|E_{n}-E_{n'}|/k_BT}\ ,\ \ \ \ \ \
\label{new-42}
\end{eqnarray}
and $\displaystyle{N_{\rm d}=\int\limits_{-{\cal L}_0/2}^{{\cal L}_0/2} d\xi\,\rho_{\rm d}(\xi)}$ represents the total number of defects in a single quantum-well structure.
For $k_BT\sim\hbar\Omega_0<\Delta E_{n,k_\|}$ (i.e., insignificant single-phonon emission regime) and ${\cal D}_{n',n}(T)\ll {\cal T}_n(k_\|,T)$ (i.e., neglecting much weaker interlevel non-radiative transitions within a defect), 
Equation\ \eqref{new-42} further leads to a statistically-averaged defect-relaxation time $\tau^{\rm d}_{\rm rel}(T,u_0,E^*)$ for all captured electrons
inside a defect, yielding

\begin{eqnarray}
\nonumber
\frac{1}{\tau^{\rm d}_{\rm rel}(T,u_0,E^*)}&=&\frac{2}{Z_{\rm eff}(T,E^*)}\sum_{n=1}^{\infty}\,\frac{n^2{\cal P}_n}{\tau^{\rm d}_n(T,u_0, E^*)}\approx\frac{2}{Z_{\rm eff}(T,E^*)}\sum_{n=1}^{\infty}\,n^2{\cal P}_n\,\\
\nonumber
&\times&\frac{2}{\rho_0(u_0){\cal A}}
\sum_{{\bf k}_\|}\,{\cal T}_n(k_\|,T)\left[f_0(\varepsilon_{k_\|}-u_0)\left(1-\texttt{e}^{-\Delta E_{n,k_{\|}}/k_BT}\right)+\texttt{e}^{-\Delta E_{n,k_{\|}}/k_BT}\right]\\
\nonumber
&\approx&\frac{4}{Z_{\rm eff}(T,E^*)}\sum_{n=1}^{\infty}\,n^2f_0(E_n-E^*)\,\frac{1}{\rho_0(u_0){\cal A}}\sum_{{\bf k}_\|}\,{\cal T}_n(k_\|,T)\\
&\times&\left\{f_0(\varepsilon_{k_\|}-u_0)+\left[1-f_0(\varepsilon_{k_\|}-u_0)\right]\,\texttt{e}^{-\Delta E_{n,k_{\|}}/k_BT}\right\}\ .
\label{eqn-39}
\end{eqnarray}
As a result, we obtain the non-equilibrium occupation function ${\cal P}_n(T,E^*,u_0)$ 
for the $n$th energy level of a defect, given by

\begin{eqnarray}
\nonumber
{\cal P}_n(T,u_0,E^*)&\approx&f_0(E_n-E^*)\\
&-&\tau^{\rm d}_{\rm rel}(T,u_0,E^*)
\left[1-f_0(E_n-E^*)\right]\frac{2}{\rho_0(u_0){\cal A}}\sum_{{\bf k}_\|}\,
f_0(\varepsilon_{k_\|}-u_0)\,{\cal T}_n(k_\|,T)\ ,
\label{eqn-add}
\end{eqnarray}
which can be employed for calculating any average physical quantities associated with a defect.
Meanwhile, using the calculated ${\cal T}_n(k_\|,T)$ and $f_0(E_n-E^*)$, we obtain
the statistically-averaged capture rate $\displaystyle{\gamma_n^{({\rm c})}(T,u_0)}$ for subband electrons by the $n$th energy level of a defect, that is,

\begin{equation}
\gamma_n^{({\rm c})}(T,u_0)=\frac{2}{\rho_0(u_0){\cal A}}
\sum_{{\bf k}_\|}\,\,{\cal T}_n(k_\|,T)\,f_0(\varepsilon_{k_\|}-u_0)\ .
\label{eqn-40}
\end{equation}
Again, this gives rise to a statistically-averaged capture rate $\gamma^{\rm d}_{\rm cap}(T,u_0,E^*)$ by a defect, i.e.,

\begin{eqnarray}
\nonumber
\gamma^{\rm d}_{\rm cap}(T,u_0,E^*)&=&\frac{1}{Z_{\rm eff}(T,E^*)}\sum_{n=1}^\infty\,2n^2\,\gamma_n^{({\rm c})}(T,u_0)\left(1-P_n\right)\\
&\approx&\frac{1}{Z_{\rm eff}(T,E^*)}\sum_{n=1}^\infty\,2n^2\left[1-f_0(E_n-E^*)\right]\frac{2}{\rho_0(u_0){\cal A}}
\sum_{{\bf k}_\|}\,\,{\cal T}_n(k_\|,T)\,f_0(\varepsilon_{k_\|}-u_0)\ .\ \ \ \ \ \ 
\label{eqn-41}
\end{eqnarray}
In a similar way, we also find  a statistically-averaged escape rate $\gamma^{\rm d}_{\rm esc}(T,u_0,E^*)$ by a defect, given by

\begin{eqnarray}
\nonumber
\gamma^{\rm d}_{\rm esc}(T,u_0,E^*)&\approx&\frac{1}{Z_{\rm eff}(T,E^*)}\sum_{n=1}^\infty\,2n^2f_0(E_n-E^*)\,\texttt{e}^{-\Delta E_{n,k_\|}/k_BT}\\
&\times&\frac{2}{\rho_0(u_0){\cal A}}
\sum_{{\bf k}_\|}\,\,{\cal T}_n(k_\|,T)\left[1-f_0(\varepsilon_{k_\|}-u_0)\right]\ ,
\label{eqn-41p}
\end{eqnarray}
where $\gamma^{\rm d}_{\rm esc}(T,u_0,E^*)<\gamma^{\rm d}_{\rm cap}(T,u_0,E^*)$ if $\Delta E_{n,k_\|}> k_BT$.
Finally, from Eq.\,\eqref{eqn-add} we get the density $\rho_{\rm cap}(T,u_0,E^*)$ for total captured subband electrons by all defects in a quantum well, yielding

\begin{eqnarray}
\nonumber
&&\rho_{\rm cap}(T,u_0,E^*)=N_{\rm d}\left\{\tau^{\rm d}_{\rm rel}(T,u_0,E^*)\left[\gamma^{\rm d}_{\rm cap}(T,u_0,E^*)-\gamma^{\rm d}_{\rm esc}(T,u_0,E^*)\right]\right\}\rho_0(u_0)\\
\nonumber
&=&N_{\rm d}\,\tau^{\rm d}_{\rm rel}(T,u_0,E^*)\,\frac{2}{{\cal A}}\sum_{{\bf k}_\|}\,\frac{1}{Z_{\rm eff}(T,E^*)}\sum_{n=1}^\infty\,2n^2
{\cal T}_n(k_\|,T)\\
&\times&
\left\{f_0(\varepsilon_{k_\|}-u_0)\left[1-f_0(E_n-E^*)\left(1-\texttt{e}^{-\Delta E_{n,k_\|}/k_BT}\,\right)\right]
-f_0(E_n-E^*)\,\texttt{e}^{-\Delta E_{n,k_\|}/k_BT}\right\}\ .\ \ \ \ \ \ \ \
\label{new-5}
\end{eqnarray}
In combination with Eqs.\,\eqref{eqn-41} and \eqref{new-5}, the charge-conservation law in  Eq.\,\eqref{eqn-25} turns into

\begin{eqnarray}
\nonumber
n_{\rm dop}&\equiv&\rho_0(u_0)+\rho_{\rm cap}(T,u_0,E^*)=\frac{2}{{\cal A}}\sum_{{\bf k}_\|}\,\left(f_0(\varepsilon_{k_\|}-u_0)
+\frac{N_{\rm d}\tau^{\rm d}_{\rm rel}(T,u_0,E^*)}{Z_{\rm eff}(T,E^*)}\,\sum_{n=1}^\infty\,2n^2
{\cal T}_n(k_\|,T)\,\right.\\
&\times&
\left.\left\{f_0(\varepsilon_{k_\|}-u_0)\left[1-f_0(E_n-E^*)\left(1-\texttt{e}^{-\Delta E_{n,k_\|}/k_BT}\,\right)\right]
-f_0(E_n-E^*)\,\texttt{e}^{-\Delta E_{n,k_\|}/k_BT}\right\}\right)\ ,
\label{eqn-44}
\end{eqnarray}
from which the chemical potential $u_0$ can be obtained as the root of Eq.\,\eqref{eqn-44} for any fixed values of $n_{\rm dop}$ and $T$, as well as for $E^*$ determined from Eq.\,\eqref{c-9}.

\section{Numerical Results and Discussions}
\label{sec-6}

In our numerical calculations, we have used parameters in Table\ \ref{tab-1}. The other parameters are set by:
${\cal L}_0=2\ell_0+a=3\ell_0$, $\rho_{\rm L}=3\times 10^{6}\,$cm$^{-1}$, $\rho_{\rm R}=2.5\times 10^{6}\,$cm$^{-1}$, $\rho_{\rm W}=1.5\times 10^{6}\,$cm$^{-1}$, $\Delta\rho=1.0\times 10^{6}\,$cm$^{-1}$, and $1\leq\kappa_0\leq 10$. Moreover, the units of energy and wave number of electrons are given by $k_F=\sqrt{2\pi n_0}=5.6\times 10^{7}\,$m$^{-1}$ and $E_F=\hbar^2 k_F^2/2m^*=1.8\,$meV, respectively, 
for selected areal density $n_0=5\times 10^{14}\,$m$^{-2}$. 
The rest of parameters, such as $N$ and $\kappa_0$ will be given directly in the figure captions. 
\medskip

\begin{figure}
\centering
\includegraphics[width=0.55\textwidth]{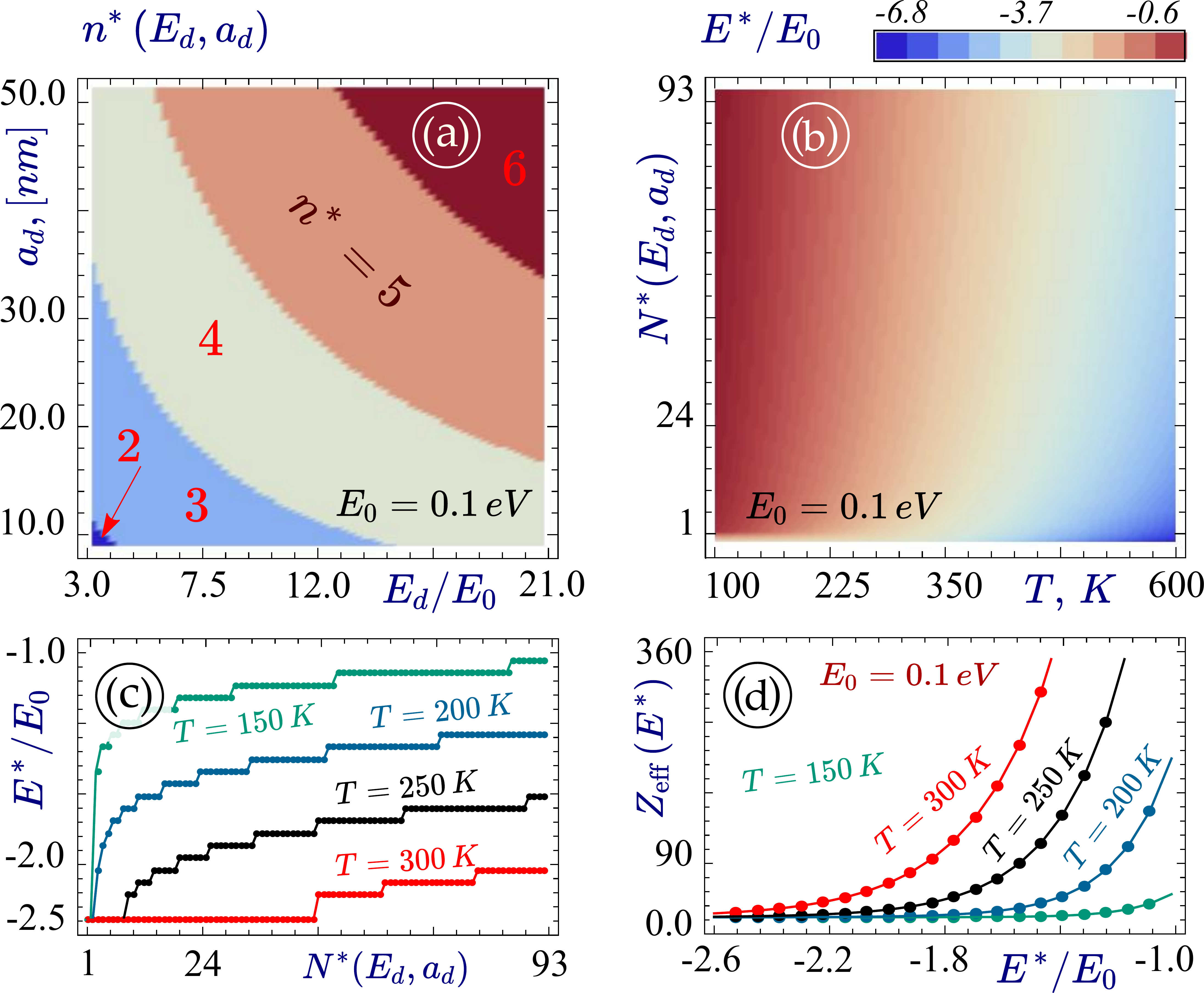}
\caption{(a) Density plot for the number of occupied levels $n^*(E_{\rm d},a_{\rm d})$, calculated from Eq.\,\eqref{c-9}, as functions of void binding energy $E_{\rm d}$ and binding radius $a_{\rm d}$. 
(b) Density plot for the chemical potential $E^*(T,N^*)$ of defect-captured electrons, calculated from Eq.\,\eqref{fermi}, 
as functions of temperature $T$ and total number of captured electrons $N^*$.
(c) Plots of $E^*(T,N^*)$ as a function of $N^*$ at different temperatures $T=150,\,200,\,250,\,300\,$K.
(d) Logarithm plots of $Z_{\rm eff}(T,E^*)$, calculated from Eq.\,\eqref{c-add}, as a function of $E^*$ at different temperatures $T=150,\,200,\,250,\,300\,$K.}
\label{fig2}
\end{figure}

Figure\ \ref{fig2}$(a)$ shows the density plot of number of occupied levels $n^*(E_{\rm d},a_{\rm d})$ from Eq.\,\eqref{c-9} 
for defect captured electrons as functions of void binding energy $E_{\rm d}$ and binding radius $a_{\rm d}$, 
where separated belts with $n^*=2,\,3,\,\dots\,6$ are seen and $n^*$ goes up with either $E_{\rm d}$ or $r_{\rm d}$.
The enhancements of $n^*$ by $E^*$ and $a_{\rm d}$ are associated with higher releasing energy from a defect-capturing process and weakened Coulomb repulsion between captured electrons. 
Figure\ \ref{fig2}$(b)$ presents the chemical potential $E^*(T,N^*)$ from Eq.\,\eqref{fermi} as functions of total number $N^*(E_{\rm d},a_{\rm d})$ for captured electrons and temperature $T$, 
where $E^*$ is found decreasing with $T$ due to smearing of Fermi surface for captured electrons but increasing with $N^*$ due to more captured electrons. 
To display specific features clearly, we also present $E^*(T,N^*)$ as a function of $N^*$ in Fig.\,\ref{fig2}$(c)$ with several values of $T$, 
where a stepwise enhancement of $E^*$ by reducing $T$, as well as by enlarging $N^*$, show up clearly in Fig.\,\ref{fig2}$(c)$. 
In Fig.\,\ref{fig2}$(d)$, we plot the calculated $E^*$ dependence for effective charge number $Z^*_{\rm eff}(T,Z^*)$ from Eq.\,\eqref{c-add} at various temperatures, 
in which the exponential increase of $Z^*_{\rm eff}(T,Z^*)$ with $E^*$ appears as expected, 
that is further supplemented by enhanced $Z^*_{\rm eff}(T,Z^*)$ with $T$ due to thermal excitation of captured electrons to higher-energy levels with much higher orbital degeneracy.
\medskip

\begin{figure}
\centering
\includegraphics[width=0.55\textwidth]{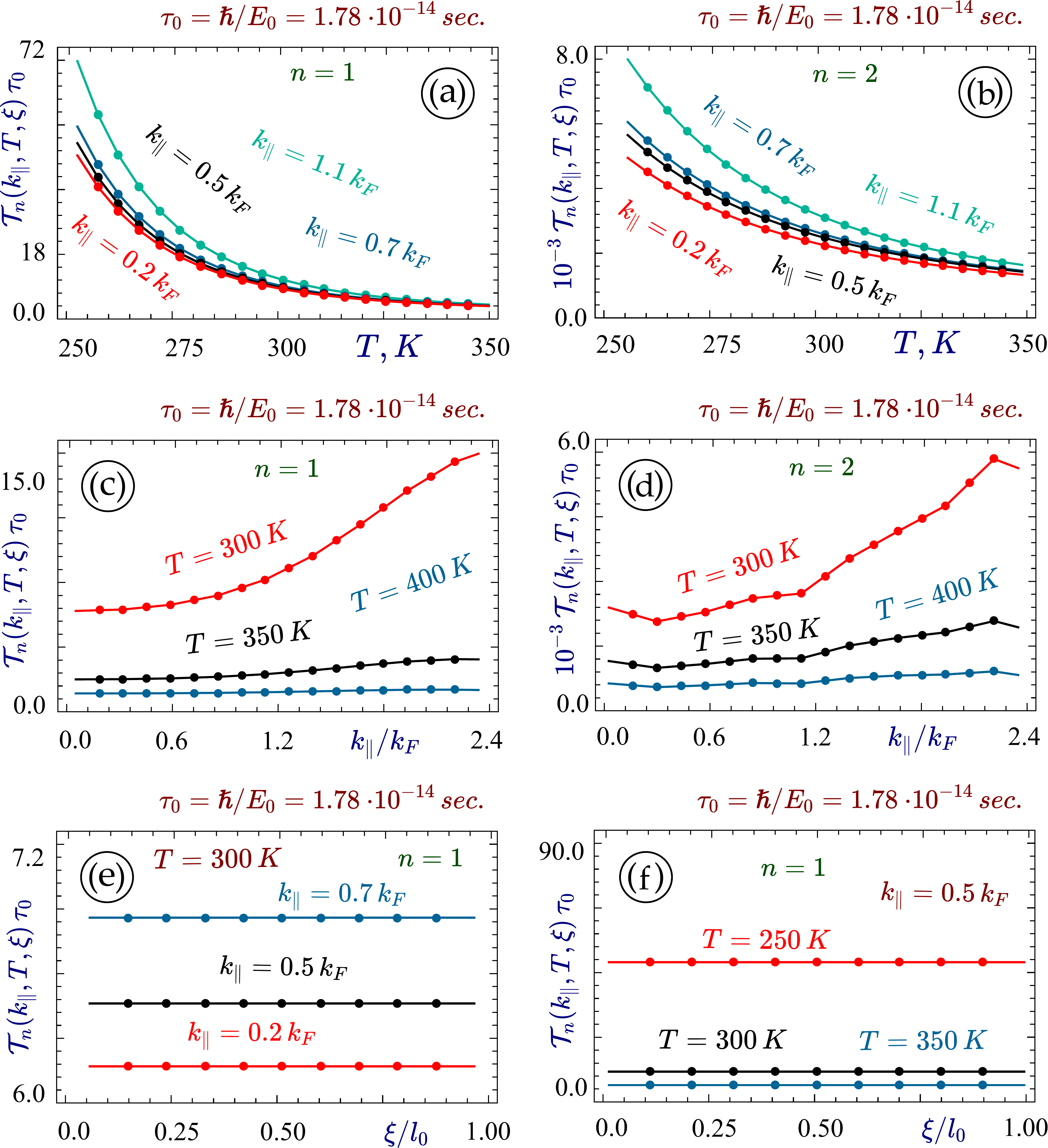}
\caption{Calculated transition rates ${\cal T}_n(k_\|,T,\xi)$ from Eq.\,\eqref{eqn-35} for $n=1,\,2$ as functions of temperature $T$ in ($a$)-($b$) and wave number $k_\|$ in ($c$)-($d$) at $\xi=0$ as well as of defect position $\xi$ in ($e$)-($f$), where $2\ell_0=2\sqrt{\hbar/m^*\omega_0}$ represents the width of a quantum well. Note that the magnitude for ${\cal T}_n(k_\|,T,\xi)$ in ($b$) and ($d$) 
has been amplified by a factor of $10^3$.}
\label{fig3}
\end{figure}

From Fig.\,\ref{fig3}$(a)$, we find that the transition rate ${\cal T}_n(k_\|,T,\xi)$ to the $n=1$ ground state of defect, 
given by Eq.\,\eqref{eqn-35}, decreases with $T$ in this high-$T$ regime (i.e., $N_0(\omega_0)\gg 1$ and $\eta_0\propto n_0(\Omega_0)\gg 1$) mainly due to reduction from the 
exponential factor ${\cal T}_n(k_\|,T,\xi)\propto\exp(\Delta E_{n,k_\|}/2k_BT)$, due to detailed balance with escape from defect,  
as $T$ goes up. Meanwhile, ${\cal T}_n(k_\|,T,\xi)$ becomes slightly 
higher for larger values of $k_\|$, which results from the minor increase of $\Delta E_{n,k_\|}$ in the same exponential factor $\exp(\Delta E_{n,k_\|}/2k_BT)$. In addition, ${\cal T}_n(k_\|,T,\xi)$ in Fig.\,\ref{fig3}$(b)$ drops three-order of magnitude for the $n=2$ excited state of defect because of significantly decreasing $\Delta E_{n,k_\|}$ with $n$. On the other hand, for ${\cal T}_n(k_\|,T,\xi)$ as functions of $k_\|$, as shown in Figs.\,\ref{fig3}$(c)$ and \ref{fig3}$(d)$, 
we observe a rise of the transition rate with $k_\|$ at $T=300\,$K due to increased $\Delta E_{n,k_\|}$ in the exponential factor $\sim\exp(\Delta E_{n,k_\|}/2k_BT)$, 
but such an enhancement is flattened out as $T$ goes up to $400\,$K due to fact of $\exp(\Delta E_{n,k_\|}/2k_BT)\to 1$ in this case. Similarly, switching from the $n=1$ ground state to $n=2$ 
excited state reduces $\Delta E_{n,k_\|}$, and then pushes down the rate ${\cal T}_n(k_\|,T,\xi)$ by three orders of magnitude 
in Fig.\,\ref{fig3}$(d)$. Finally, as displayed in Figs.\,\ref{fig3}$(e)$ and \ref{fig3}$(f)$, 
it is revealed that the transition rate ${\cal T}_n(k_\|,T,\xi)$ becomes largely independent of $\xi$ within the range of $|\xi|\leq\ell_0$. This feature is attributed to the fact that the void binding radius $a_{\rm d}$ becomes much larger than the quantum-well 
width $2\ell_0=2\sqrt{\hbar/m^*\omega_0}$ in this case, and therefore, $\left|\langle\psi_{n,\ell,m}(\mbox{\boldmath$r$}_\|,z-\xi)\vert
U_{\rm ep}(\mbox{\boldmath$r$}_\|,z)\vert\phi_{k_\|}(\mbox{\boldmath$r$}_\|,z)\rangle\right|^2$ in Eq.\,\eqref{eqn-35} is independent of $\xi$ as can be verified from Eqs.\,\eqref{eqn-36} and \eqref{eqn-37}. However, such $\xi$ dependence in ${\cal T}_n(k_\|,T,\xi)$ will be fully recovered once the condition $a_{\rm d}\leq\ell_0$ or $|\xi-\ell_0|>a_{\rm d}$ is satisfied.
\medskip

\begin{figure}	
\centering
\includegraphics[width=0.65\textwidth]{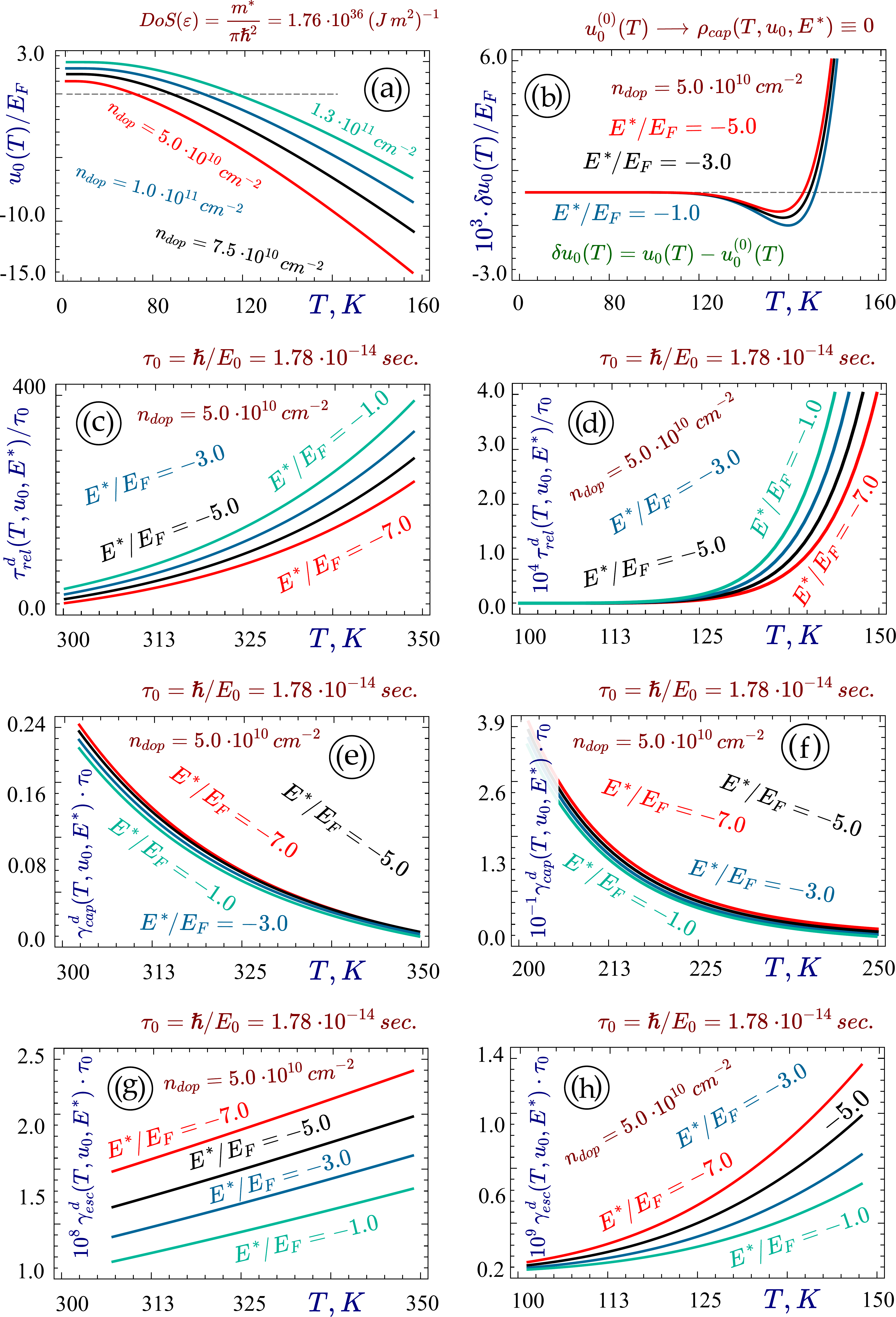}
\caption{Temperature dependence of calculated chemical potential $u_0(T)$ from Eqs.\,\eqref{new-5}-\eqref{eqn-44} in ($a$) 
and its variation $\delta u_0(T)\equiv u_0(T)-u^{(0)}_0(T)$ in ($b$) 
with $u_0^{(0)}(T)$ as the chemical potential corresponding to $\rho_{\rm cap}(T,u_0,E^*)=0$, along with   
the defect-relaxation time $\tau_{\rm rel}^{\rm d}(T,u_0,E^*)$ from Eq.\,\eqref{eqn-39} in ($c$), ($d$),
the defect-capture rate $\gamma_{\rm cap}^{\rm d}(T,u_0,E^*)$ from Eq.\,\eqref{eqn-41} in ($e$), ($f$),
and the defect-escape rate $\gamma_{\rm esc}^{\rm d}(T,u_0,E^*)$ from Eq.\,\eqref{eqn-41p} in ($g$), ($h$),
for different values of $n_{\rm dop}$ and $E^*$. Here, the vertical scales in panels ($b$), ($d$), ($g$), ($h$) are amplified by factors of $10^3,\,10^4,\, 10^8,\, 10^9$, respectively, while 
the vertical scale in panel ($f$) is compressed by a factor of $10^{-1}$. Moreover, the dashed lines in ($a$) and ($b$) indicate the zero scale.}
\label{fig4}
\end{figure}

In Fig.\,\ref{fig4}$(a)$, by including the defect-capture contributions $\rho_{\rm cap}(T,u_0,E^*)$ obtained from Eq.\,\eqref{new-5}, we show the self-consistently calculated 
chemical potential $u_0(T)$ as a function of temperature $T$ for different values of doping density $n_{\rm dop}$. In this case, $u_0(T)$ is seen 
increasing with $n_{\rm dop}$ for each given $T$ but decreasing from positive to negative for fixed $n_{\rm dop}$ as $T$ goes up. 
Due to the presence of $\rho_{\rm cap}(T,u_0,E^*)$ in Eq.\,\eqref{eqn-44}, 
we find from Fig.\,\ref{fig4}$(b)$ that the variation $\delta u_0(T)\equiv u_0(T)-u^{(0)}_0(T)$ of the chemical potential depends on $T$ in a non-monotonic way 
and exhibits an initial rolling down followed by a sharp rising around its negative minimum. This unique corner-like feature reveals 
the fundamental role played by $\rho_{\rm cap}(T,u_0,E^*)$  
in Eq.\,\eqref{eqn-44} for given $n_{\rm dop}$ and is attributed to an interplay between the increasing defect-relaxation time $\tau^{\rm d}_{\rm rel}(T,u_0,E^*)$ ($\sim 10^{-12}\,$s)
and the decreasing defect-capture rate $\gamma^{\rm d}_{\rm cap}(T,u_0,E^*)$ ($\sim 10^{15}\,$s$^{-1}$) as functions of $T$, as shown by Eq.\,\eqref{new-5}. 
The increasing $\tau^{\rm d}_{\rm rel}(T,u_0,E^*)$ in Eq.\,\eqref{eqn-39} and the decreasing defect-capture rate $\gamma^{\rm d}_{\rm cap}(T,u_0,E^*)$ Eq.\,\eqref{eqn-41} 
as functions of $T$ are presented in Figs.\,\ref{fig4}$(c)$, \ref{fig4}$(d)$ and Figs.\,\ref{fig4}$(e)$, \ref{fig4}$(f)$, respectively. 
Both thermally-reduced $1/\tau^{\rm d}_{\rm rel}(T,u_0,E^*)$ and $\gamma^{\rm d}_{\rm cap}(T,u_0,E^*)$ are all due to decreasing transition rate ${\cal T}_n(k_\|,T)$ with $T$.
Meanwhile, we find $\delta u_0(T)$ increases slightly with $|E^*|$ around the corner. 
However, $\tau^{\rm d}_{\rm rel}(T,u_0,E^*)$ decreases significantly with $|E^*|$ for high $T$, in contrast with the slight increase of $\gamma^{\rm d}_{\rm cap}(T,u_0,E^*)$ with $T$. 
Furthermore, the defect-escape rate $\gamma^{\rm d}_{\rm esc}(T,u_0,E^*)$ ($\sim 10^{6}\,$s$^{-1}$) in Figs.\,\ref{fig4}$(g)$ and \ref{fig4}$(h)$ is also enhanced by increasing $T$, 
which results from the weakening of its exponential factor $\sim\exp(-\Delta E_{n,k_\|}/k_BT)$ in Eq.\,\eqref{eqn-41p} with increasing $T$, and simultaneously, 
it increases with growing $|E^*|$ at high $T$.
\medskip

\begin{figure}
\centering
\includegraphics[width=0.65\textwidth]{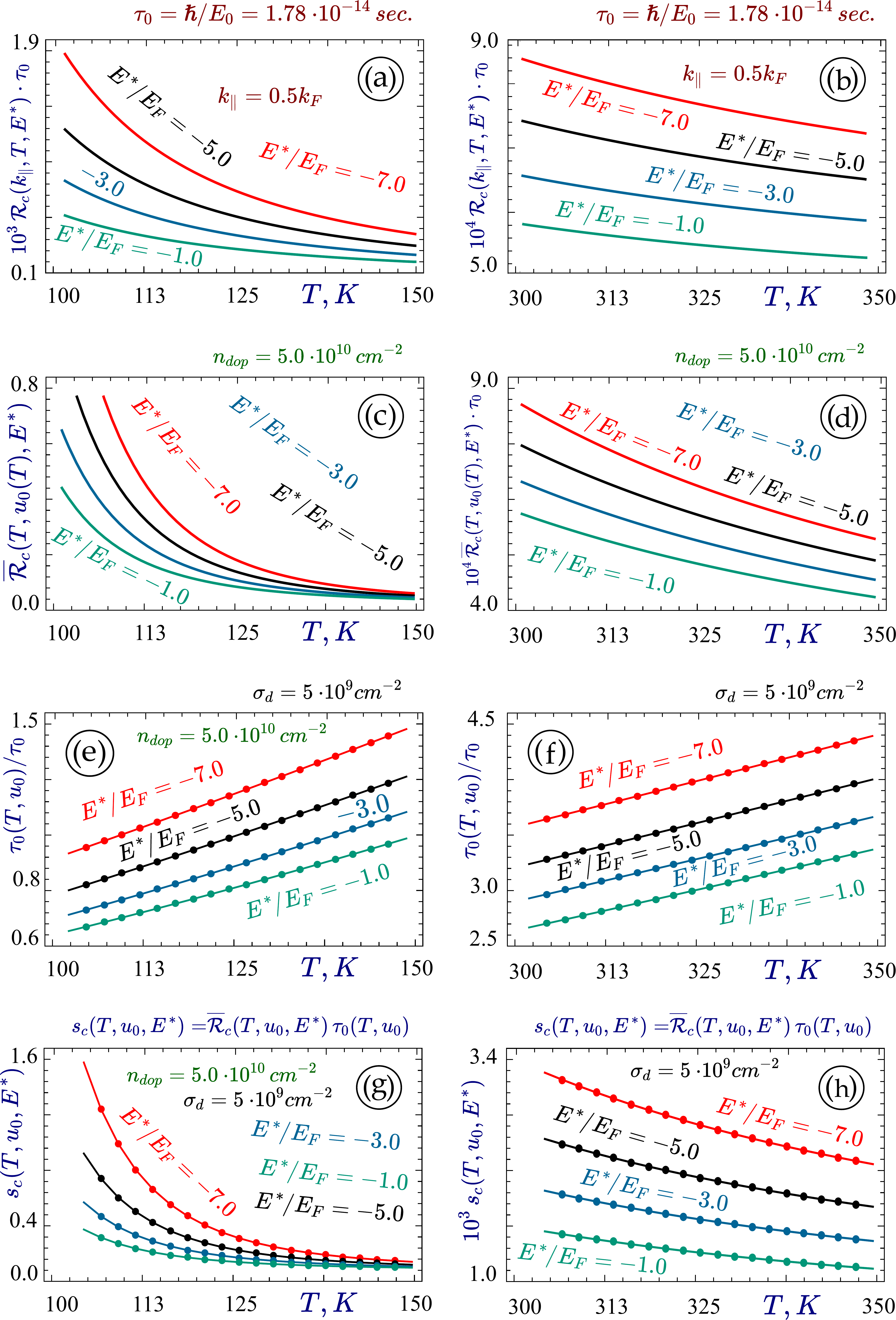}
\caption{Temperature dependence of calculated capture rate ${\cal R}_{\rm c}(k_\|,T,E^*)$ of Bloch electrons from Eq.\,\eqref{new-36} in ($a$), ($b$), 
statistically-averaged capture rate $\overline{{\cal R}}_{\rm c}(T,u_0,E^*)$ from Eq.\,\eqref{new-24} in ($c$), ($d$), 
and statistically-averaged energy-relaxation time $\tau_0(T,u_0)$ from Eqs.\,\eqref{eqn-20} and \eqref{a-1} in ($e$), ($f$),
as well as current-suppression factor $s_{\rm c}(T,u_0,E^*)=\overline{{\cal R}}_{\rm c}(T,u_0,E^*)\tau_0(T,u_0)$ in ($g$), ($h$),
for various values of $E^*$. Here, the vertical scales in panels ($a$), ($b$), ($d$), ($h$) are amplified by factors of $10^3,\,10^4,\,10^4,\,10^3$, respectively.}
\label{fig5}
\end{figure}

By applying the calculated defect transition rate ${\cal T}_n(k_\|,T)$ to Eq.\,\eqref{new-36}, we display in Figs.\,\ref{fig5}$(a)$ and \ref{fig5}$(b)$  
the capture rate ${\cal R}_{\rm c}(k_\|,T,E^*)$ ($\sim 10^{11}\,$s$^{-1}$) of Bloch electrons as a function of temperature $T$ for different values of $E^*$. 
In this case, ${\cal R}_{\rm c}(k_\|,T,E^*)$ is found decreasing with $T$ but is enhanced by $|E^*|$ at each given $T$. 
Meanwhile, ${\cal R}_{\rm c}(k_\|,T,E^*)$ increases with $k_\|$ due to ${\cal T}_n(k_\|,T,\xi)$ as demonstrated in Figs.\,\ref{fig3}$(c$) and \ref{fig3}$(d$).
Additionally, the statistically-averaged capture rate $\overline{{\cal R}}_{\rm c}(T,u_0,E^*)$ ($\sim 10^{14}\,$s$^{-1}$) 
calculated from Eq.\,\eqref{new-24} is shown in Figs.\,\ref{fig5}$(c)$ and \ref{fig5}$(d)$ 
as a function of $T$ for various values of $E^*$, and it depends on $u_0(T)$ and reduces with increasing $T$. 
Here, we observe that $\overline{{\cal R}}_{\rm c}(T,u_0,E^*)$ becomes much bigger at low $T$ but 
decreases with $T$ much faster in comparison with ${\cal R}_{\rm c}(k_\|,T,E^*)$ presented in Figs.\,\ref{fig5}$(a)$ and \ref{fig5}$(b)$.
Moreover, the very short statistically-averaged energy-relaxation time $\tau_0(T,u_0)$ ($\sim 10^{-14}\,$s)
calculated from Eq.\,\eqref{eqn-20} and Eq.\,\eqref{a-1} appears in Figs.\,\ref{fig5}$(e)$ and \ref{fig5}$(f)$ as a function of $T$, 
which is revealed increasing with $T$ as well as enhanced by $E^*$ for fixed $T$.
Furthermore, the dimensionless current-suppression factor $s_{\rm c}(T,u_0,E^*)=\overline{{\cal R}}_{\rm c}(T,u_0,E^*)\tau_0(T,u_0)$ ($\sim 1.6$ around $100\,$K for more than $60\%$ capturing)
in Figs.\,\ref{fig5}$(g)$ and \ref{fig5}$(h)$, which is initially introduced in Eq.\,\eqref{eqn-24},   
is found to be a fast decreasing function of $T$ in the low-$T$ regime and increases with $|E^*|$. 
However, we expect a very large current-suppression factor as $T$ becomes much lower than $100\,$K. For $T>150\,$K, on the other hand, we expect 
$s_{\rm c}(T,u_0,E^*)\ll 1$ for a weak-capturing process and it becomes negligible above $300\,$K. 
\medskip

\begin{figure}
\centering
\includegraphics[width=0.95\textwidth]{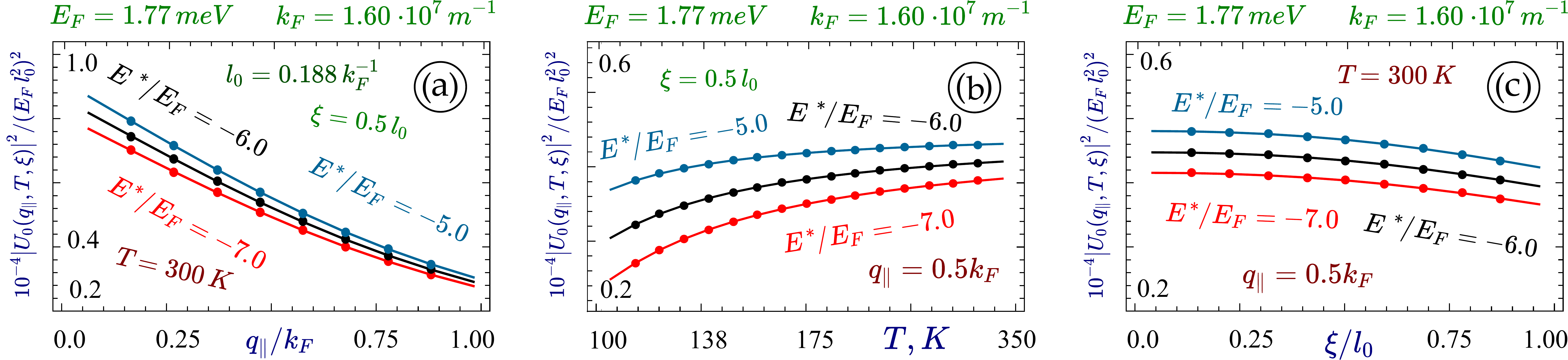}
\caption{Calculated electron-defect interaction $|\overline{U}_{\rm d}(q_\|,T)|^2$ from Eq.\,\eqref{eqn-8} for various values of $E^*$ 
as a function of $q_\|$ at $T=300\,$K and $\xi/\ell_0=0.5$ in ($a$), a function of $T$ with $q_\|/k_F=0.5$ and $\xi/\ell_0=0.5$ in ($b$), as well as  
a function of $\xi$ at $q_\|/k_F=0.5$ and $T=300\,$K in ($c$). Here, the vertical scales of $|\overline{U}_{\rm d}(q_\|,T)|^2$ in all three panels 
are compressed by a factor of $10^{-4}$.}
\label{fig6}
\end{figure}

The electron-defect interactions $|\overline{U}_{\rm d}(q_\|,T,\xi)|^2$ obtained from Eq.\,\eqref{eqn-8} with $\xi/\ell_0=0.5$ are shown in Figs.\,\ref{fig6}$(a)$ and \ref{fig6}$(b)$   
as functions of $q_\|$ and $T$, respectively. From Fig.\,\ref{fig6}$(a)$, we first observe that $|\overline{U}_{\rm d}(q_\|,T,\xi)|^2$ goes down with $q_\|$ due to reduced Coulomb interaction 
between electrons and defects as a result of enhanced $\Delta_0(q_\|,\phi)\propto q_\|$ factor in Eq.\,\eqref{eqn-8}. 
At the same time, $|\overline{U}_{\rm d}(q_\|,T,\xi)|^2$ also reduces with increaseing $|E^*|$ for each given $q_\|$ 
because of the presence of $Z_{\rm eff}(T,E^*)$ factor in Eq.\,\eqref{eqn-8}, as seen from Fig.\,\ref{fig2}$(d)$.
Moreover, from Fig.\,\ref{fig6}$(b)$ we find that $|\overline{U}_{\rm d}(q_\|,T,\xi)|^2$ increases with $T$, but decreases with $|E^*|$ for fixed $T$.  
This property can also be attributed to $Z_{\rm eff}(T,E^*)$ factor appearing in Eq.\,\eqref{eqn-8}, as demonstrated by Fig.\,\ref{fig2}$(d)$. 
Finally, $|\overline{U}_{\rm d}(q_\|,T,\xi)|^2$ in Eq.\,\eqref{eqn-8} is shown in Figs.\,\ref{fig6}$(c)$ for $q_\|/k_F=0.5$ and $T=300\,$K, and 
from that we know $|\overline{U}_{\rm d}(q_\|,T,\xi)|^2$ attains its maximum at $\xi=0$, 
and then drops with increasing $\xi$ values away from the well center at $z=0$. This fact results from the maximized ground-state wave function $|{\cal F}_0(z)|$ at $z=0$ in Eq.\,\eqref{eqn-8}.
\medskip

\begin{figure}
\centering
\includegraphics[width=0.75\textwidth]{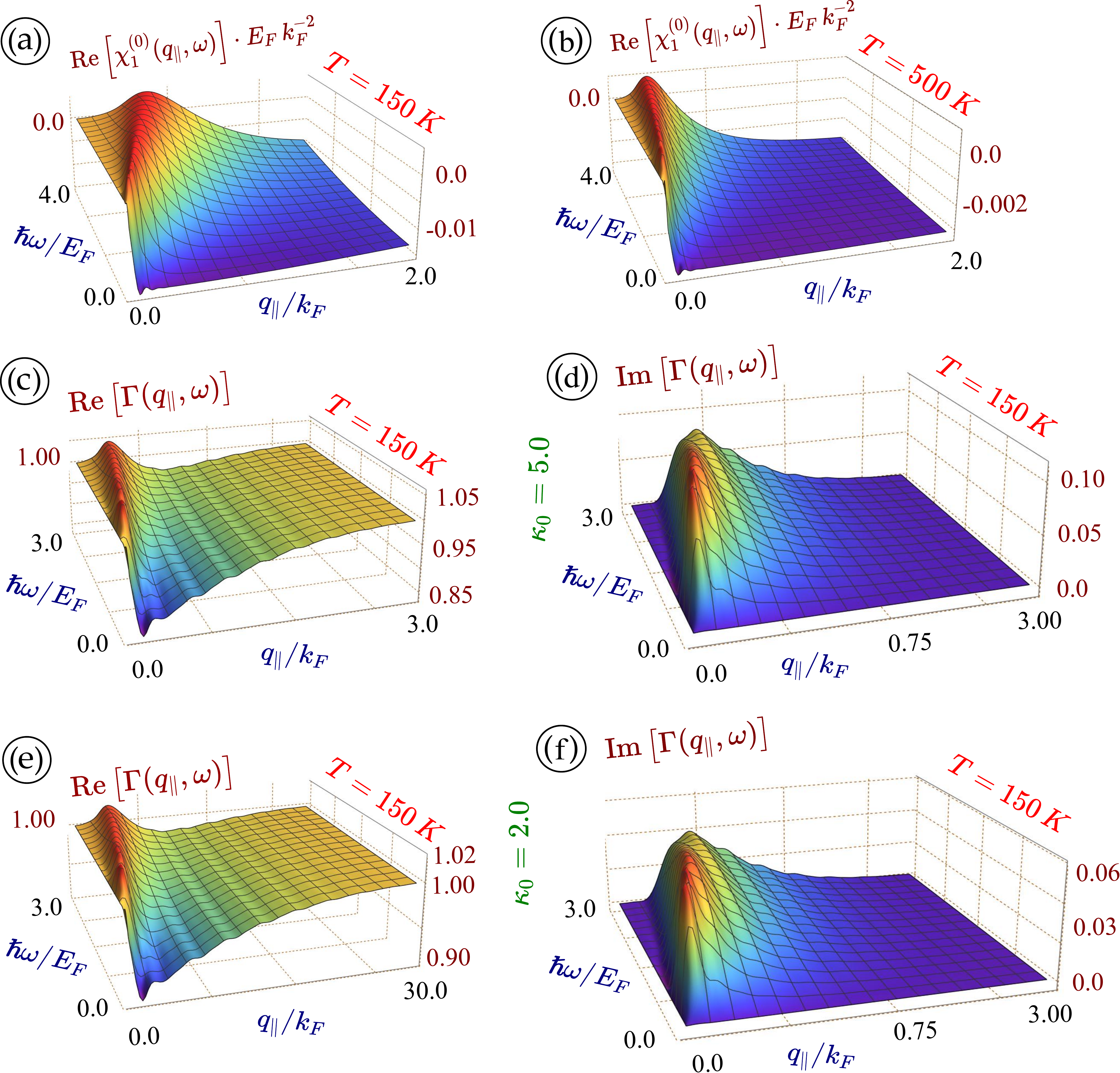}
\caption{3D plots as functions of $q_\|$ and $\omega$ 
for real ({\rm Re}) and imaginary ({\rm Im}) parts of the polarization function $\chi^{(0)}_1(q_\|,\omega)$ calculated from Eq.\,\eqref{eqn-2} at $T=150\,$K and $500\,$K in ($a$) and ($b$), respectively, 
as well as for ${\rm Re}[\Gamma(q_\|,\omega)]$ and ${\rm Im}[\Gamma(q_\|,\omega)]$ of 
defect-vertex corrections obtained from Eq.\,\eqref{eqn-9} at $T=150\,$K  with $\kappa_0=5$ in ($c$), ($d$) and $\kappa_0=2$ in ($e$), ($f$).}
\label{fig7}
\end{figure}

As seen in Fig.\,\ref{fig7}$(a)$, the bare ${\rm Re}[\chi^{(0)}_1(q_\|,\omega)]$ from Eq.\,\eqref{eqn-2} acquires a positive resonant peak as a function of $\omega$ as $q_\|/k_F\ll 1$, and 
the observed peak decays rapidly with increasing $q_\|$. Meanwhile, for fixed $q_\|$,  
this peak rolls down with decreasing $\omega$ towards a negative value reached at $\omega=0$. Moreover, the peak strength of ${\rm Re}[\chi^{(0)}_1(q_\|,\omega)]$ is greatly weakened 
by raising temperature $T$ from $150\,$K to $500\,$K, as demonstrated in Fig.\,\ref{fig7}$(b)$. 
On the other hand, we find from Eq.\,\eqref{eqn-9} that ${\rm Re}[\Gamma(q_\|,\omega)]\equiv 1$ and ${\rm Im}[\Gamma(q_\|,\omega)]\equiv 0$ if $\rho_{\rm d}(\xi)=0$ has been assumed corresponding to the 
absence of defects. However, if $\rho_{\rm d}(\xi)\,|\overline{U}_{\rm d}(q_\|,T,\xi)|^2\ll 1$ is satisfied, we have $\Gamma(q_\|,\omega)\approx 1+(2m^*/\pi^2\hbar^2)\,\chi_1^{(0)}(q_\|,\omega)
\int_{-{\cal L}_0/2}^{{\cal L}_0/2}\,d\xi\,\rho_{\rm d}(\xi)\,\left|\overline{U}_{\rm d}(q_\|,T,\xi)\right|^2$. 
As a result, ${\rm Re}[\Gamma(q_\|,\omega)]$ in Fig.\,\ref{fig7}($c$) is also expected to  display a positive peak correspondingly, and becomes either above or below unity, depending on the positive or negative sign of ${\rm Re}[\chi_1^{(0)}(q_\|,\omega)]$ as a function of $\omega$ around this resonant peak. Furthermore, we know from Fig.\,\ref{fig6}$(a)$ that 
$|\overline{U}_{\rm d}(q_\|,T,\xi)|^2\ll 1$ as $q_\|/k_F\gg 1$, and this leads to 
${\rm Re}[\Gamma(q_\|,\omega)]\to 1$ in Fig.\,\ref{fig7}$(c)$ and ${\rm Im}[\Gamma(q_\|,\omega)]\to 0$ in Fig.\,\ref{fig7}$(d)$ simultaneously for $q_\|/k_F\gg 1$. 
Finally, after $\kappa_0$ is reduced from $5.0$ to $2.0$ in Figs.\,\ref{fig7}$(e)$ and \ref{fig7}$(f)$,
the strength of previous positive resonant peak in both Figs.\,\ref{fig7}$(c)$ and \ref{fig7}$(d)$ are scaled down proportionally as expected, 
but leaving the overall shape unchanged in Figs.\,\ref{fig7}$(e)$ and \ref{fig7}$(f)$. 
\medskip

\begin{figure}
\centering
\includegraphics[width=0.65\textwidth]{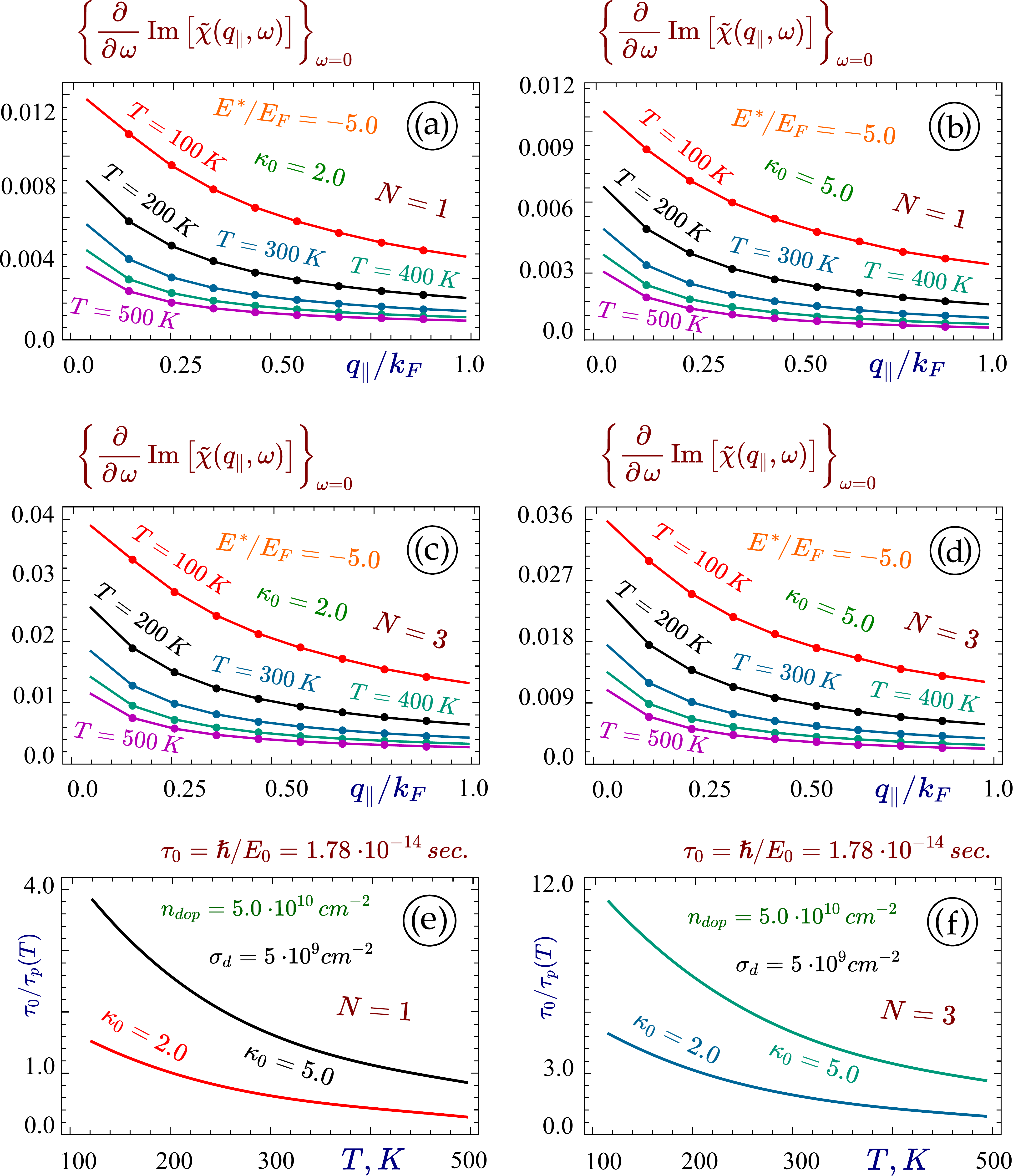}
\caption{The calculated $\omega$-derivative of imaginary parts of the polrization function $\partial{\rm Im}[\tilde{\chi}(q_\|,\omega)]/\partial\omega$ from Eqs.\,\eqref{eqn-10}, \eqref{eqn-1} and \eqref{eqn-2} at $\omega=0$ as a function of $q_\|$ with different temperatures $T$ for single ($N=1$) and triple ($N=3$) quantum wells are displayed in Figs.\,\ref{fig8}$(a)$, \ref{fig8}$(b)$ and Figs.\,\ref{fig8}$(c)$, \ref{fig8}$(d)$, respectively, where we assume $\kappa_0=2.0$ and $5.0$ in Figs.\,\ref{fig8}$(a)$, \ref{fig8}$(c)$ and in Figs.\,\ref{fig8}$(b)$, \ref{fig8}$(d)$. Meanwhile, the obtained  
inverse of momentum-relaxation times $1/\tau_p$ from Eq.\,\eqref{eqn-34} with $\kappa_0=2.0$ and $5.0$ as a function of $T$ are presented in Figs.\,\ref{fig8}$(e)$ and \ref{fig8}$(f)$ for single and triple quantum wells.}
\label{fig8}
\end{figure}

We have presented the bare polarization function $\chi_1^{(0)}(q_\|,\omega)$ in Eq.\,\eqref{eqn-2}, while its defect-vertex correction, as well as its intrawell screening correction, are given in Eqs.\,\eqref{eqn-1} and \eqref{eqn-4}, respectively, for single-well ($N=1$) system. Meanwhile, we have also included the interwell screening correction in Eq.\,\eqref{eqn-10} for multi-well system ($N>1$). From Fig.\,\ref{fig8}$(a)$, for $N=1$ we first find that $\partial{\rm Im}[\tilde{\chi}(q_\|,\omega)]/\partial\omega=\partial{\rm Im}[\chi_1(q_\|,\omega)]/\partial\omega$ at $\omega=0$ decreases with $q_\|$ from its resonance peak at $q_\|=0$ for each fixed temperature $T$ until approaching zero as $q_\|/k_F\gg 1$. interwell screening correction also reduces with increasing $T$ at the same time due to occupation of high-energy states. As $\kappa_0$ is lifted from $2.0$ to $5.0$ in Fig.\,\ref{fig8}$(b)$, 
$\partial{\rm Im}\chi_1(q_\|,\omega)]/\partial\omega$ at $\omega=0$ is further reduced because of enhanced vertex-correction $\Gamma(q_\|,\omega)$ and then screening effect $\epsilon(q_\|,\omega)$ in Eq.\,\eqref{eqn-4}, as indicated by Eq.\,\eqref{eqn-1}. For $N=3$ in Figs.\,\ref{fig8}$(c)$ and \ref{fig8}$(d)$, $\partial{\rm Im}[\tilde{\chi}(q_\|,\omega)]/\partial\omega$ at $\omega=0$ increases proportionally in this case compared to results in Figs.\,\ref{fig8}$(a)$ and \ref{fig8}$(b)$ correspondingly, which stems from the fact that $\tilde{\chi}(q_\|,\omega)\propto N\chi_1(q_\|,\omega)$. Finally, for inverse momentum-relaxation time $1/\tau_p$ presented in Figs.\,\ref{fig8}$(e)$ and \ref{fig8}$(f)$ as a function of temperature $T$, we demonstrate that it decreases with $T$ but increases with $N$, as found from Figs.\,\ref{fig8}$(a)$ and \ref{fig8}$(c)$. Here, the decrease and increase of $1/\tau_p$ with $T$ and $N$ are caused, respectively, by occupation of high-energy states and the relation $\tilde{\chi}(q_\|,\omega)\propto N$, as clearly shown in Figs.\,\ref{fig8}$(a)$-\ref{fig8}$(d)$.
Furthermore, $1/\tau_p$ is also strengthened by $\kappa_0$ as expected from the enhanced  contribution of $|U_0(q_\|,T)|^2\propto\kappa_0$ in Eq.\,\eqref{eqn-29} to $1/\tau_p$ in Eq.\,\eqref{eqn-34}.
\medskip

\begin{figure}
\centering
\includegraphics[width=0.65\textwidth]{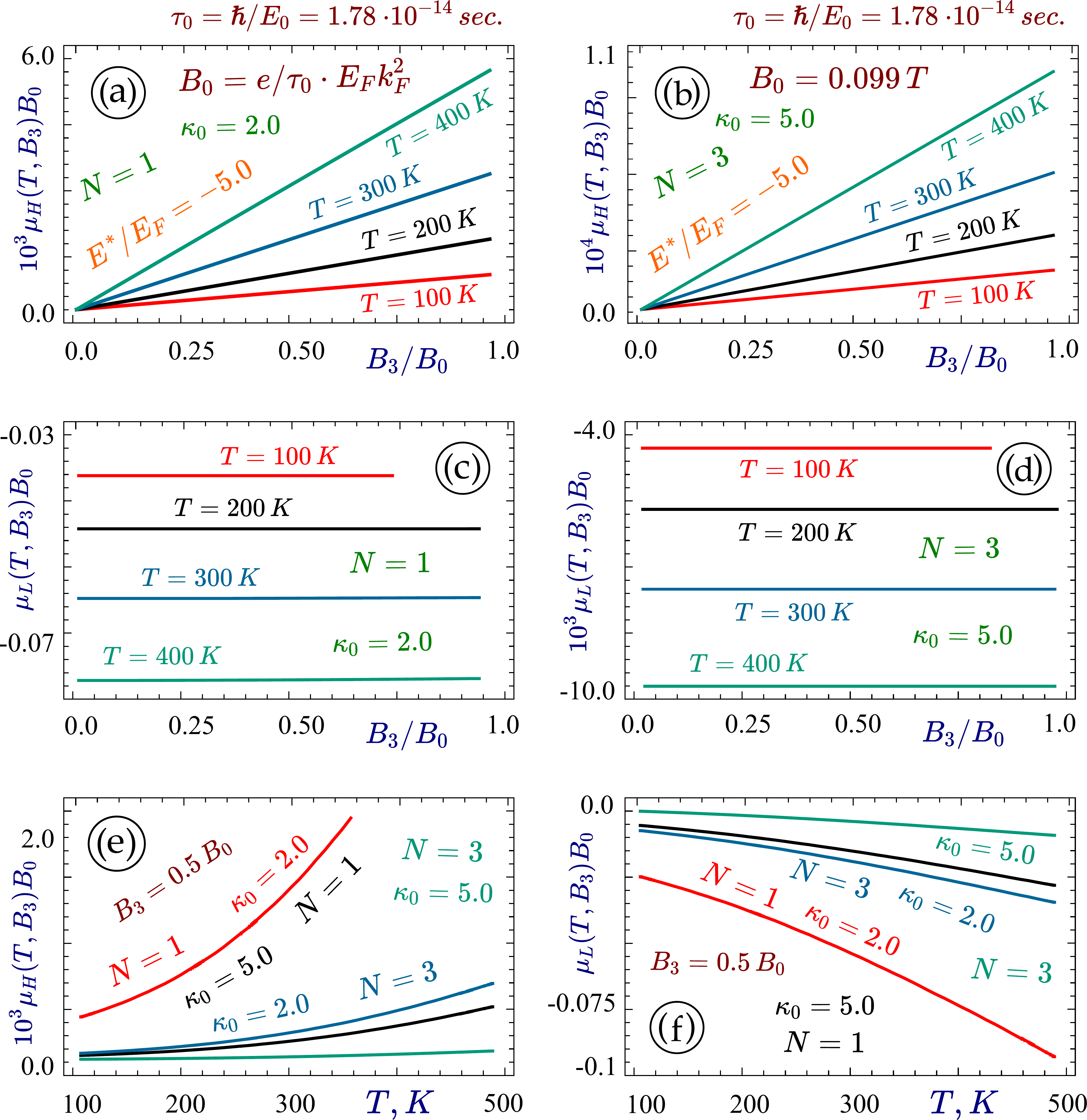}
\caption{The calculated Hall $\mu_{\rm H}(T,u_0,B_3)$ and longitudinal $\mu_{\rm L}(T,u_0,B_3)$ mobilities from Eq.\,\eqref{b-12} are displayed in ($a$), ($b$) and in ($c$), ($d$), respectively, as a function of perpendicular magnetic field $B_3$ for different temperatures $T$ with $\kappa_0=2.0$, $N=1$ in ($a$), ($c$) and with $\kappa_0=5.0$, $N=3$ in ($b$), ($d$). Moreover, $\mu_{\rm H}(T,u_0,B_3)$ and $\mu_{\rm L}(T,u_0,B_3)$ as a function of $T$ are also presented in ($e$), ($f$) for various of combinations of $N=1,\,3$ and $\kappa_0=2.0,\,5.0$. Here, factors of $10^4$ and $10^3$ have been introduced in ($b$) and in ($a$), ($d$) and ($e$).}
\label{fig9}
\end{figure}

We present the calculated Hall mobility $\mu_{\rm H}(T,u_0,B_3)$ as a function of the perpendicular magnetic field $B_3$ for $N=1,\,\kappa_0=2.0$ in Fig.\,\ref{fig9}$(a)$ as well as for $N=3,\,\kappa_0=5.0$ in Fig.\,\ref{fig9}$(b)$ with various values of temperatures $T$. As $B_3$ increases, $\mu_{\rm H}(T,u_0,B_3)$ goes up linearly with $\mu_0^2\,B_3$ in Fig.\,\ref{fig9}$(a)$ for $\mu^2_0B^2_3\ll 1$, as shown in Eq.\,\eqref{b-12}. Meanwhile, we also find $\mu_{\rm H}(T,u_0,B_3)$ increasing with $T$ for fixed $B_3$ due to decrease of $1/\tau_p$ as seen from Fig.\,\ref{fig8}$(e)$. However, the increase of $\mu_{\rm H}(T,u_0,B_3)$ with $B_3$ becomes much slower in Fig.\,\ref{fig9}$(b)$ for the case with $N=3,\,\kappa_0=5.0$, which can be attributed to the enhanced momentum-relaxation rate $1/\tau_p$ by larger values of $N$ and $\kappa_0$ as verified from Fig.\,\ref{fig8}$(f)$. In addition, we also exhibit the calculated longitudinal mobility $\mu_{\rm L}(T,u_0,B_3)$ as a function of perpendicular magnetic field $B_3$ for $N=1,\,\kappa_0=2.0$ in Fig.\,\ref{fig9}$(c)$ an for $N=3,\,\kappa_0=5.0$ in Fig.\,\ref{fig9}$(d)$ with various values $T$. In this case, however, we find $\mu_{\rm L}(T,u_0,B_3)$ in Figs.\,\ref{fig9}$(c)$ and \ref{fig9}$(d)$ become independent of $B_3$ due to $\mu^2_0B^2_3\ll 1$ in Eq.\,\eqref{b-12}.
Moreover, $|\mu_{\rm L}(T,u_0,B_3)|$ increases with either $N$ or $\kappa_0$, as demonstrated in Fig.\,\ref{fig9}$(d)$ and explained by Figs.\,\ref{fig8}$(e)$ and \ref{fig9}$(f)$. Finally, both $\mu_{\rm H}(T,u_0,B_3)$ and $\mu_{\rm L}(T,u_0,B_3)$ as functions of $T$ are given in 
Figs.\,\ref{fig9}$(e)$ and \ref{fig9}$(f)$, from which we observe that $\mu_{\rm H}(T,u_0,B_3)$ and $|\mu_{\rm L}(T,u_0,B_3)|$ increases with $T$ but decreases with either $N$ or $\kappa_0$, as known from Figs.\,\ref{fig8}$(e)$ and \ref{fig9}$(f)$.

\section{Conclusions and Remarks}
\label{sec-7}

In conclusion, we have calculated the defect corrections to the polarization and dielectric functions of Bloch electrons in quantum wells. Meanwhile,
we have also derived the first two moment equations from the semi-classical Boltzmann transport theory and applied them for investigating defect effects on magneto-transport of electrons.
By using this defect-corrected polarization function, we have further acquired analytically both the momentum-relaxation time and the mobility tensor for Bloch electrons.
Using quantum-statistical theory, we have explored defect capture and charging dynamics based on a parameterized quantum-mechanics model for defects and by employing a finite-range probability function. 
Consequently, we have obtained both the capture, escape and relaxation rates and the density of captured Bloch electrons as functions of temperature, doping density and various types of defects.
Such a microscopic-level theory is very important and applicable to many electronic systems, including accurately quantifying burst noise in transistors and blinking noise in photo-detectors. 
\medskip

In particular, by employing the energy-balance equation, we have first determined the number of occupied energy levels $N^*$ and the chemical potential $E^*$ of defects for captured Bloch electrons as a function of temperature $T$ with various values of defect binding energy and binding radius. Meanwhile, we have also studied the transition rate ${\cal T}(k_\|,T)$ for defect-capturing process as functions of Bloch-electron wave number $k_\|$ and $T$. Based on calculated $E^*$ and ${\cal T}(k_\|,T)$, we have further explored the defect energy-relaxation $\tau^d_{rel}$, capture $\gamma^d_{cap}$ and escape $\gamma^d_{esc}$ rates, in addition to the self-consistent chemical potential $u_0(T)$ of Bloch electrons, as functions of $T$ and doping density $n_{dop}$. On the other hand, we have further calculated the average Bloch-electron energy-relaxation $1/\tau_0(T)$ and momentum-relaxation $1/\tau_p(T)$ rates, as well as the Bloch-electron current suppression factor $s_c(T,u_0,E^*)$, as a function of $T$. Finally, we have investigated the magnetic-field $B_3$ and $T$ dependence of both Hall $\mu_H(T,B_3)$ and longitudinal $\mu_L(T,B_3)$ mobilities of Bloch electrons in single and triple quantum wells.

\begin{acknowledgements}
DH would like to acknowledge the financial supports from the Air Force Office of Scientific Research (AFOSR)
and the Laboratory University Collaboration Initiative (LUCI) program.
\end{acknowledgements}

\newpage
\appendix

\section{Energy-Relaxation Time}
\label{app-1}

By using the detailed-balance condition, the energy-relaxation time $\tau_{\rm d}(k_\|)$ initially introduced in Eq.\,(\ref{eqn-20}) for defects can be calculated according to

\begin{equation}
\frac{1}{\tau_{\rm d}(k_\|)}={\cal W}_{\rm in}(k_\|)+{\cal W}_{\rm out}(k_\|)=\frac{4m^*\sigma_{\rm d}}{\pi\hbar^3}\int\limits_{0}^{\pi} d\theta\,\left|U_0(2k_\||\cos\theta|)\right|^2\ ,
\label{a-1}
\end{equation}
where $\sigma_{\rm d}$ is the point-defect areal density. The scattering-in rate in Eq.\,\eqref{a-1} for electrons in the final ${\mbox{\boldmath$k$}}_\|$-state is

\begin{eqnarray}
\nonumber
{\cal W}_{\rm in}(k_\|)&=&\frac{2\pi\sigma_{\rm d}}{\hbar{\cal A}}\,\sum_{{\bf q}_\|}\,\left|U_0(q_\|)\right|^2
\left[f_{{\bf k}_\|-{\bf q}_\|}\,\delta(\varepsilon_{{\bf k}_\|}-\varepsilon_{{\bf k}_\|-{\bf q}_\|})+f_{{\bf k}_\|+{\bf q}_\|}\,
\delta(\varepsilon_{{\bf k}_\|}-\varepsilon_{{\bf k}_\|+{\bf q}_\|})\right]\\
&=&\left(\frac{4m^*\sigma_{\rm d}}{\pi\hbar^3}\right)f_0(\varepsilon_{k_\|})\int\limits_{0}^{\pi} d\theta\,\left|U_0(2k_\||\cos\theta|)\right|^2\ ,
\label{a-2}
\end{eqnarray}
and the scattering-out rate for electrons in the initial ${\mbox{\boldmath$k$}}_\|$-state is

\begin{eqnarray}
\nonumber
{\cal W}_{\rm out}(k_\|)&=&\frac{2\pi\sigma_{\rm d}}{\hbar{\cal A}}\,\sum_{{\bf q}_\|}\,\left|U_0(q_\|)\right|^2\\
\nonumber
&\times&\left[(1-f_{{\bf k}_\|+{\bf q}_\|})\,\delta(\varepsilon_{{\bf k}_\|+{\bf q}_\|}-\varepsilon_{{\bf k}_\|})+(1-f_{{\bf k}_\|-{\bf q}_\|})\,
\delta(\varepsilon_{{\bf k}_\|-{\bf q}_\|}-\varepsilon_{{\bf k}_\|})\right]\\
&=&\left(\frac{4m^*\sigma_{\rm d}}{\pi\hbar^3}\right)\,[1-f_0(\varepsilon_{k_\|})]\int\limits_{0}^{\pi} d\theta\,\left|U_0(2k_\||\cos\theta|)\right|^2\ .
\label{a-3}
\end{eqnarray}
For simplicity, we have introduced the notations, i.e., $f_{{\bf k}_\|}\equiv f_0(\varepsilon_{k_\|}-u_0)$.
We further assume that electrons are remotely doped, temperature $T$ is low, and areal electron density $\rho_0$ is low.
As a result, we can neglect the electron scatterings with ionized impurities, lattice phonons and other electrons,
and retain the electron scattering only with charged defects in the system.
\medskip

For the electron-defect scattering in a quantum well, its interaction introduced in Eqs.\,\eqref{a-2} and \eqref{a-3} takes the form

\begin{eqnarray}
\nonumber
&&\left|U_0(q_\|)\right|^2=\left(\frac{a_0}{a_{\rm d}}\right)^2\,\left[\frac{Z_{\rm eff}(T,E^*)\,e^2}{2\epsilon_0\epsilon_{\rm d}}\right]^2\,
\frac{\texttt{e}^{-q^2_\|\Lambda^2_\|/2}}{(q_\|+q_{\rm s})^2}\\
&\times&
\int\limits_{-{\cal L}_0/2}^{{\cal L}_0/2} d\xi\,\rho_{\rm d}(\xi)
\left[\int\limits_{0}^{\infty} dz\int\limits_{-\infty}^\infty dz'\,\left|{\cal F}_{0}(z)\right|^2\left(\texttt{e}^{-q_\||z-z'|}+\texttt{e}^{-q_\||z+z'|}\right){\cal Q}_1(q_\|,z'-\xi)\right]^2,\ \ \ \ \
\label{a-4}
\end{eqnarray}
where $Z_{\rm eff}(T,E^*)$ is the charge number of defects, $a_{\rm d}$ and $a_0=5\,$\AA\ are the void and point-vacancy binding radii,  
$\epsilon_{\rm d}$ the dielectric constant of the host semiconductor,
$1/q_{\rm s}$ the inverse of a static screening length,
$\Lambda_\|$ is the in-plane correlation length for randomly-distributed defects,
the partial form factor ${\cal Q}_1(q_\|,\xi)$ is given by Eq.\,\eqref{c-8},
and $\rho_{\rm d}(\xi)$ stands for a one-dimensional density distribution of defects.

\section{Mobility Tensor}
\label{app-2}

From the force-balance equation in Eq.\,(\ref{eqn-21}), by using $\displaystyle{\tensor{\mbox{\boldmath${\tau}$}_{p}}^{-1}=\frac{1}{\tau_j}\,\delta_{ij}}$ for simplicity,
we get the following two inhomogeneous linear equations for $\mbox{\boldmath$v$}_d=\{v_1,v_2,0\}$

\begin{equation}
\left(1+q\tau_1r_{12}B_3\,\right)v_1-q\tau_1r_{11}B_3\,v_2
=q\tau_1\left(r_{11}E_1+r_{12}E_2\right)\ ,
\label{b-1}
\end{equation}

\begin{equation}
q\tau_2r_{22}B_3\,v_1+\left(1-q\tau_2r_{21}B_3\right)v_2
=q\tau_2\left(r_{21}E_1+r_{22}E_2\right)\ ,
\label{b-2}
\end{equation}
where we have used the notations ${\mbox{\boldmath$B$}}=\{0,0,B_3\}$, ${\mbox{\boldmath$E$}}=\{E_1,E_2,0\}$, and $\tensor{\mbox{\boldmath${\cal M}$}}^{-1}=\{r_{ij}\}$ for $i,j=1,\,2$.
By defining the coefficient matrix $\tensor{\mbox{\boldmath${\cal C}$}}$ for the above linear equations, i.e.,

\begin{equation}
\tensor{\mbox{\boldmath${\cal C}$}}=
\left[\begin{array}{cc}
1+q\tau_1r_{12}B_3 & -q\tau_1r_{11}B_3\\
q\tau_2r_{22}B_3 & 1-q\tau_2r_{21}B_3
\end{array}\right]\ ,
\label{b-3}
\end{equation}
as well as the source vector ${\mbox{\boldmath$s$}}$, given by

\begin{equation}
{\mbox{\boldmath$s$}}=\left[\begin{array}{c}
q\tau_1(r_{11}E_1+r_{12}E_2)\\
q\tau_2(r_{21}E_1+r_{22}E_2)
\end{array}\right]\ ,
\label{b-4}
\end{equation}
we can reduce the above linear equations to a matrix equation $\tensor{\mbox{\boldmath${\cal C}$}}\cdot{\mbox{\boldmath$v$}}_d={\mbox{\boldmath$s$}}$ with the formal solution
$\mbox{\boldmath$v$}_d=\tensor{\mbox{\boldmath${\cal C}$}}^{-1}\cdot{\mbox{\boldmath$s$}}$. Explicitly, we find the solution $\mbox{\boldmath$v$}_d=\{v_1,\,v_2,\,0\}$
for $j=1,\,2$ by

\begin{equation}
v_j=\frac{Det\{\tensor{\mbox{\boldmath${\Delta}$}}_j\}}{Det\{\tensor{\mbox{\boldmath${\cal C}$}}\}}\ ,
\label{b-5}
\end{equation}
where

\footnotesize
\begin{equation}
\tensor{\mbox{\boldmath${\Delta}$}}_1=
\left[\begin{array}{cc}
q\tau_1(r_{11}E_1+r_{12}E_2) & -q\tau_1r_{11}B_3\\
q\tau_2(r_{21}E_1+r_{22}E_2) & 1-q\tau_2r_{21}B_3
\end{array}\right]\ ,
\label{b-6}
\end{equation}

\begin{equation}
\tensor{\mbox{\boldmath${\Delta}$}}_2=
\left[\begin{array}{cc}
1+q\tau_1r_{12}B_3 & q\tau_1(r_{11}E_1+r_{12}E_2)\\
q\tau_2r_{22}B_3 & q\tau_2(r_{21}E_1+r_{22}E_2)
\end{array}\right]\ .
\label{b-7}
\end{equation}
\medskip

\normalsize
Moreover, by assuming $r_{ij}=0$ for $i\neq j$, $\displaystyle{r_{jj}=\frac{1}{m_j^\ast}}$ and introducing the notation $\displaystyle{\mu_j=\frac{q\tau_j}{m_j^\ast}}$, we find

\begin{equation}
\tensor{\mbox{\boldmath${\cal C}$}}=
\left[\begin{array}{cc}
1 & -\mu_1B_3\\
\mu_2B_3 & 1
\end{array}\right]\ ,
\label{b-8}
\end{equation}

\begin{equation}
\tensor{\mbox{\boldmath${\Delta}$}}_1=
\left[\begin{array}{cc}
\mu_1E_1 & -\mu_1B_3\\
\mu_2E_2 & 1
\end{array}\right]\ ,
\label{b-9}
\end{equation}

\begin{equation}
\tensor{\mbox{\boldmath${\Delta}$}}_2=
\left[\begin{array}{cc}
1 & \mu_1E_1\\
\mu_2B_3 & \mu_2E_2
\end{array}\right]\ ,
\label{b-10}
\end{equation}
and
\[
Det\{\tensor{\mbox{\boldmath${\cal C}$}}\}=1+\mu_1\mu_2\,B_3^2\ ,
\]

\[
Det\{\tensor{\mbox{\boldmath${\Delta}$}}_1\}=\mu_1E_1+\mu_1\mu_2\,B_3E_2\ ,
\]

\[
Det\{\tensor{\mbox{\boldmath${\Delta}$}}_2\}=\mu_2E_2-\mu_2\mu_1\,B_3E_1\ .
\]
From the above results, we finally arrive at

\begin{equation}
\tensor{\mbox{\boldmath${\mu}$}}(B_3)=\frac{1}{1+\mu_1\mu_2B_3^2}\,
\left[\begin{array}{cc}
\mu_1 & \mu_1\mu_2B_3\\
-\mu_2\mu_1B_3 & \mu_2
\end{array}\right]\ .
\label{b-11}
\end{equation}
\medskip

If we further assume $q=-e$, $m_1^\ast=m_2^\ast=m^\ast$ and $\tau_1=\tau_2=\tau_p$, we obtain
$Det\{\tensor{\mbox{\boldmath${\cal C}$}}\}=1+\mu^2_0B_3^2$,
$Det\{\tensor{\mbox{\boldmath${\Delta}$}}_1\}=-\mu_0E_1+\mu_0^2B_3E_2$, and
$Det\{\tensor{\mbox{\boldmath${\Delta}$}}_2\}=-\mu_0E_2-\mu_0^2B_3E_1$,
where $\displaystyle{\mu_0=\frac{e\tau_p}{m^\ast}}$. This gives rise to

\begin{equation}
\tensor{\mbox{\boldmath${\mu}$}}({\mbox{\boldmath$B$}})=-\frac{\mu_0}{1+\mu_0^2B_3^2}\,
\left[\begin{array}{cc}
1 & -\mu_0B_3\\
\mu_0B_3 & 1
\end{array}\right]\ .
\label{b-12}
\end{equation}

\section{Defect Effective Potential and Occupation}
\label{app-3}

The Schr\"odinger equation for electrons bounded to a defect can be formally written as\,\cite{dft-3}

\begin{equation}
-\frac{\hbar^2}{2m^\ast}\,\nabla_{{\bf r}}^2\,\psi_{\nu}(\mbox{\boldmath$r$})+[V_L(\mbox{\boldmath$r$})-V_D(\mbox{\boldmath$r$})]\,\psi_{\nu}(\mbox{\boldmath$r$})=E_{\nu}\,\psi_{\nu}(\mbox{\boldmath$r$})\ ,
\label{c-1}
\end{equation}
where $V_L(\mbox{\boldmath$r$})$ is the full crystal lattice-potential energy,
$V_D(\mbox{\boldmath$r$})\approx V_L(\mbox{\boldmath$r$})-U_A(\mbox{\boldmath$r$})$ represents the deformed lattice-potential energy by a defect,
$U_A(\mbox{\boldmath$r$})$ is the local atomic-potential energy for a point vacancy,
$m^\ast$ is the effective mass of Bloch electrons,
and $\psi_{\nu}({\bf r})$ and $E_{\nu}$ represent the eigen-function and eigen-energy of defect-bound electrons, respectively.
The accurate calculation of $V_D(\mbox{\boldmath$r$})$ can be done with first-principles density-functional theory for a point defect.\,\cite{dft-1,dft-2,dft-3}
\medskip

For simplicity, we assume $\displaystyle{U_A(\mbox{\boldmath$r$})=-\frac{Z_de^2}{4\pi\epsilon_0\epsilon_{\rm s}r}}$ for a spherical defect with $Z_d$
approximately as a net charge number and $\epsilon_{\rm s}$ being the dielectric constant of a host material.
Moreover, we assume the defect-state energy levels $\displaystyle{E_n=-\frac{E_{\rm d}}{n^2}}$
for the principal quantum number $n=1,\,2,\,3,\,\cdots$, $E_{\rm d}$ for the binding energy,
and $a_{\rm d}$ for the void radius, as well as the corresponding wave functions

\begin{eqnarray}
\nonumber
\psi_{\nu}(\mbox{\boldmath$r$})&\equiv&\psi_{n,\ell,m}(\mbox{\boldmath$r$})=R_{n,\ell}(r^*)\,\overline{Y}_{\ell,m}(\cos\theta)\,\frac{\texttt{e}^{im\phi}}{\sqrt{2\pi}}\ ,\\
\nonumber
R_{n,\ell}(r^*)&=&N^{(1)}_{n\ell}\,\exp\left(-\frac{r^*}{n}\right)\,\left(\frac{2r^*}{n}\right)^\ell\,L^{\{2\ell+1\}}_{n+\ell}\left(\frac{2r^*}{n}\right)\ ,\\
\overline{Y}_{\ell,m}(\cos\theta)&=&(-1)^m\,N^{(2)}_{\ell m}\,P^m_\ell(\cos\theta)\ ,
\label{c-2}
\end{eqnarray}
where $\mbox{\boldmath$r$}\equiv\{\mbox{\boldmath$r$}_\|,z\}$ is a position vector,
the angular-momentum quantum number $\ell=0,\,1,\,2,\,\cdots,\,n-1$,
the magnetic quantum number $m=0,\,\pm 1,\,\pm 2,\,\cdots,\,\pm\ell$, $r=\sqrt{r_\|^2+z^2}$, $\displaystyle{\cos\theta=\frac{z}{r}}$, $r^*=r/a_{\rm d}$,
and the two normalization factors are given by

\begin{eqnarray}
\nonumber
N^{(1)}_{n\ell}&=&\sqrt{\left(\frac{2}{na_{\rm d}}\right)^3\frac{(n-\ell-1)!}{2n[(n+\ell)!]^3}}\ ,\\
\nonumber
N^{(2)}_{\ell,m}&=&\sqrt{\frac{(2\ell+1)\,(\ell-|m|)!}{2\,(\ell+|m|)!}}\ .
\end{eqnarray}
Additionally, $L_n^{\{\alpha\}}(x)$ in Eq.\,\eqref{c-2} are the generalized Laguerre polynomials, and $P_\ell^m(x)$ are the associated Legendre polynomials.
\medskip

Using the obtained bound-electron energy levels $E_n=-E_{\rm d}/n^2$ and wave functions $\psi_{n,\ell,m}(\mbox{\boldmath$r$})$ in Eq.\,\eqref{c-2}, the effective potential energy
$U_{\rm eff}(\mbox{\boldmath$r$})$ of a charged defect takes the form

\begin{equation}
U_{\rm eff}(\mbox{\boldmath$r$})=\frac{2e^2}{4\pi\epsilon_0\epsilon_{\rm s}}\,\sum_{n=1}^{\infty}\,f_0(E_n-E^*)
\sum_{\ell,m}\,\int d^3\mbox{\boldmath$r$}'\,\frac{\left|\psi_{n,\ell,m}(\mbox{\boldmath$r$}')\right|^2}{|\mbox{\boldmath$r$}-\mbox{\boldmath$r$}'|}\ ,
\label{c-3}
\end{equation}
where $f_0(x)=\left[1+\exp\left(x/k_BT\right)\right]^{-1}$ is the Fermi function,
$E^*\in\max\{-E_{\rm d},\,0\}$ is the Fermi energy for bound electrons inside a defect, and $T$ is the lattice temperature.
If we simply replace $\left|\psi_{n,\ell,m}(\mbox{\boldmath$r$}')\right|^2$ in Eq.\,\eqref{c-3} by a function $\delta(\mbox{\boldmath$r$}')$, $U_{\rm eff}(\mbox{\boldmath$r$})$ reduces to the well-known Coulomb potential energy
$\displaystyle{U_{\rm c}(\mbox{\boldmath$r$})=\frac{Z_{\rm eff}(T,E^*)\,e^2}{4\pi\epsilon_0\epsilon_{\rm s}r}}$ with

\begin{equation}
Z_{\rm eff}(T,E^*)=2\sum_{n=1}^\infty\,n^2f_0(E_n-E^*)
\label{c-add}
\end{equation}
as an effective charge number,
where $k_BT$ much less than the level separation $\Delta E_n$ is assumed in the last step, and
both the orbital and spin degeneracies of bound electrons are included.
\medskip

By introducing a two-dimensional (2D) Fourier transform for a quantum-well system, we have

\begin{equation}
\frac{1}{|\mbox{\boldmath$r$}-\mbox{\boldmath$r$}'|}=\frac{2\pi}{{\cal A}}\sum_{{\bf q}'_\|}\,\exp[i\mbox{\boldmath$q$}'_\|\cdot(\mbox{\boldmath$r$}_\|-\mbox{\boldmath$r$}'_\|)]\,\frac{\texttt{e}^{-q'_\||z-z'|}}{q'_\|}\ ,
\label{c-4}
\end{equation}
where ${\cal A}$ is the cross-sectional area of the quantum well. Then, from Eq.\,\eqref{c-3} we get the 2D Fourier transformed $U_{\rm eff}(\mbox{\boldmath$q$}_\|,z)$, given by

\begin{eqnarray}
\nonumber
U_{\rm eff}(\mbox{\boldmath$q$}_\|,z)&=&\int d^2\mbox{\boldmath$r$}_\|\,U_{\rm eff}(\mbox{\boldmath$r$}_\|,z)\,\exp(-i\mbox{\boldmath$q$}_\|\cdot\mbox{\boldmath$r$}_\|)\\
&=&\frac{e^2}{\epsilon_0\epsilon_{\rm s}q_\|}\,\sum_{n=1}^{\infty}\,f_0(E_n-E^*)
\int dz'\,\texttt{e}^{-q_\||z-z'|}\,{\cal Q}_{n}(\mbox{\boldmath$q$}_\|,z')\ ,\ \ \ \
\label{c-5}
\end{eqnarray}
where the partial form factor ${\cal Q}_{n}(\mbox{\boldmath$q$}_\|,z')$ is defined as

\begin{equation}
{\cal Q}_{n}(\mbox{\boldmath$q$}_\|,z')=\int d^2\mbox{\boldmath$r$}'_\|\,\exp(-i\mbox{\boldmath$q$}_\|\cdot\mbox{\boldmath$r$}'_\|)\,\sum_{\ell,m}\,\left|\psi_{n,\ell,m}(\mbox{\boldmath$r$}'_\|,z')\right|^2\ .
\label{c-6}
\end{equation}
By approximating

\[
\left|\psi_{n,\ell,m}(\mbox{\boldmath$r$}_\|,z)\right|^2\approx \frac{1}{\pi}\,\left(\frac{1}{a_{\rm d}}\right)^3\,\exp\left[-(2/a_{\rm d})\,(r_\|^2+z^2)^{1/2}\right]
\]
for the lowest $n=1$ eigenstate, we arrive at an explicit expression

\begin{equation}
U_{\rm eff}(\mbox{\boldmath$q$}_\|,z)\approx\frac{Z_{\rm eff}(T,E^*)\,e^2}{2\epsilon_0\epsilon_{\rm s}(q_\|+q_s)}\int\limits_{-\infty}^\infty dz'\,
\texttt{e}^{-q_\||z-z'|}\,\texttt{e}^{-q_\|^2\Lambda^2_\|/4}\,{\cal Q}_{1}(q_\|,z')\ ,
\label{c-7}
\end{equation}
where

\begin{equation}
{\cal Q}_{1}(q_\|,z)=2\left(\frac{1}{a_{\rm d}}\right)^3\int\limits_0^\infty dr_\|\,r_\|J_0(q_\|r_\|)\,\texttt{e}^{-(2/a_{\rm d})\,\sqrt{r_\|^2+z^2}}\ ,
\label{c-8}
\end{equation}
$J_0(x)$ is the first-kind Bessel function of order zero, and its contribution becomes negligible 
for either $r_\|/a_{\rm d}\gg 1$ or $|z|/a_{\rm d}\gg 1$.
\medskip

Physically, we still need to determine the value of $E^*$ for thermal-equilibrium distribution of trapped electrons inside a defect.
Considering the increased Coulomb repulsion by filling more and more electrons into a defect,
we first introduce the following {\em quantum-mechanical energy-balance equation}, i.e.,

\begin{equation}
\frac{e^2}{4\pi\epsilon_0\epsilon_{\rm s}}\,\sum_{s=1}^{n^*}\,\sum_{\ell=0}^{s-1}\,\frac{2\ell+1}{|\bar{\cal R}_{n^*+1}-\bar{r}_{s,\ell}|}
-\left[\frac{\hbar\omega_0}{2}+\frac{E_{\rm d}}{(n^*+1)^2}\right]=0\ ,
\label{c-9}
\end{equation}
where $n^*(E_{\rm d},a_{\rm d})\in\max\{1,2,3,\cdots\}$, which represents the index of the topmost occupied energy level of the defect, is the root of Eq.\,\eqref{c-9}, 
$\displaystyle{\frac{\hbar\omega_0}{2}}$ the lowest subband edge, 
and the quantum-mechanically averaged orbital radii of trapped electrons are 

\begin{eqnarray}
\bar{r}_{s,\ell}&=&\int\limits_0^\infty\,\left|R_{s,\ell}(r)\right|^2r^3dr=s^2a_{\rm d}\,\left[1+\frac{1}{2}\left(1-\frac{\ell(\ell+1)}{s^2}\right)\right]\ ,\\
\bar{\cal R}_{n^*+1}&=&\frac{1}{n^*+1}\,\sum\limits_{\ell=0}^{n^*}\,\bar{r}_{n^*+1,\ell}\ .
\label{c-10}
\end{eqnarray}
For this obtained index $n^*(E_{\rm d},a_{\rm d})$, the total number $N^*(E_{\rm d},a_{\rm d})$ for captured electrons is calculated as
$\displaystyle{N^*(E_{\rm d},a_{\rm d})=2\sum\limits_{s=1}^{n^*}\,s^2}$ including spin degeneracy. 
Next, for a given value $N^*(E_{\rm d},a_{\rm d})$ resulting from the root of Eq.\,\eqref{c-9}, we further put forward the following {\em particle-number conservation equation} 
to include the thermal effect at temperature $T$, yielding

\begin{equation}
2\,\sum_{n=1}^{\infty}\,\frac{n^2}{1+\exp[(E_n-E^*)/k_{\rm B}T]}-N^*(E_{\rm d},a_{\rm d})=0\ ,
\label{fermi}
\end{equation}
where the temperature-dependent $E^*(T,N^*)$ is the root of Eq.\,\eqref{fermi} and represents the ``chemical potential'' for captured electrons in a defect.

\clearpage
\begin{table}[htbp]
	\centering
	\caption{\bf Parameters Used for Numerical Calculations}
	\begin{tabular}{llll}
		\hline
		Parmeter\ \ \ \ 			&	Description\ \ \ \ \ \ \ \ \ \ \ \ \ \ \ \ \ \ \ \ \ \ \ \ \ \ \ \ 	& 	Value\ \ \ \	& 	Units 	    \\
		\hline
		$\hbar\omega_0$				&	level separation		        	& 75		&	meV					\\
		$m^*/m_0$                   &   effective mass                      & 0.067     &                       \\  
		$a_{\rm d}$		            &	void binding radius			        & 10  		&   nm				    \\
		$a_0$		                &	point vacancy binding radius	    & 5  		&   \AA				    \\
		$E_{\rm d}$ 				&	defect binding energy           	& 300		&	meV					\\
        $\sigma_{\rm d}$            &   point-defect areal density          & 5         &   $10^{9}\,$cm$^{-2}$\\
		$T$                         &   temperature                         & 300       &   K                   \\ 
		$\hbar\Omega_0$             &   phonon energy                       & 36        &   meV                 \\
		$\epsilon_{\rm s}$          &   dielectric constant                 & 11.9      &                       \\ 
		$\Lambda_\|$                &   correlation length                  & 10        &   nm                  \\      
		$\gamma_0$                  &   level broadening                    & 1         &   meV                 \\
		$n_{\rm dop}$               &   doping density                      & 5         &   $10^{10}\,$cm$^{-2}$\\   
		$\epsilon_{\rm L} $         &   static dielectric constant          & 12.9      &                       \\   
		$\epsilon_{\rm H}$          &   optic dielectric constant           & 10.89     &                       \\   
        ${\cal S}_0$                &   Huang-Rhys factor                   & 1.1       &                       \\
		\hline
	\end{tabular}
	\label{tab-1}
\end{table}

\end{document}